\documentclass[aps,prl,preprint,nopacs,superscriptaddress]{revtex4-1}

\usepackage{graphicx}
\usepackage{verbatim}
\usepackage{mathrsfs}
\usepackage{gensymb}
\usepackage{color}
\usepackage{colortbl}
\usepackage{ulem}
\usepackage{amsmath,amsfonts,amssymb}

\usepackage{braket}

\newcommand{\bs}[1]{{\boldsymbol{#1}}}

\newcommand{\blue}[1]{{\textcolor{blue}{#1}}}
\definecolor{mygray}{gray}{.9}
\definecolor{mypink}{rgb}{.99,.91,.95}
\definecolor{mycay}{rgb}{.3,.75,.93}
\newcommand{\bee}{\begin{equation}}
\newcommand{\ee}{\end{equation}}
\def\3{2.8in}    
\def\2{2.5in}
\def\4{3.0in}\def \beq {\begin{equation}}
\def \eeq {\end{equation}}
\makeatletter

    \def\CT@@do@color{%
      \global\let\CT@do@color\relax
            \@tempdima\wd\z@
            \advance\@tempdima\@tempdimb
            \advance\@tempdima\@tempdimc
    \advance\@tempdimb\tabcolsep
    \advance\@tempdimc\tabcolsep
    \advance\@tempdima2\tabcolsep
            \kern-\@tempdimb
            \leaders\vrule
                    \hskip\@tempdima\@plus  1fill
            \kern-\@tempdimc
            \hskip-\wd\z@ \@plus -1fill }
    \makeatother

\begin{document}

\title{Topological quantum properties of chiral crystals}

\author{Guoqing~Chang$^*$\footnote[0]{*These authors contributed equally to this work.}}
\affiliation{Laboratory for Topological Quantum Matter and Advanced Spectroscopy (B7), Department of Physics, Princeton University, Princeton, New Jersey 08544, USA}
\affiliation{Centre for Advanced 2D Materials and Graphene Research Centre, National University of Singapore, Singapore;}
 \affiliation{Department of Physics, National University of Singapore, 2 Science Drive 3, 117546, Singapore}
\affiliation{Institute of Physics, Academia Sinica, Taipei 11529, Taiwan}

\author{Benjamin~J.~Wieder$^*$}
\affiliation{Department of Physics, Princeton University, Princeton, New Jersey 08544, USA}
\affiliation{Nordita, Center for Quantum Materials, KTH Royal Institute of Technology and Stockholm University, Roslagstullsbacken 23, SE-106 91 Stockholm, Sweden}
\affiliation{Department of Physics and Astronomy, University of Pennsylvania, Philadelphia, Pennsylvania 19104, USA}

\author{Frank~Schindler$^*$}
\affiliation{Department of Physics, University of Zurich, Winterthurerstrasse 190, 8057 Zurich, Switzerland}

\author{Daniel~S.~Sanchez}
\affiliation{Laboratory for Topological Quantum Matter and Advanced Spectroscopy (B7), Department of Physics, Princeton University, Princeton, New Jersey 08544, USA}

\author{Ilya~Belopolski}
\affiliation{Laboratory for Topological Quantum Matter and Advanced Spectroscopy (B7), Department of Physics, Princeton University, Princeton, New Jersey 08544, USA}

\author{Shin-Ming~Huang}
\affiliation{Department of Physics, National Sun Yat-Sen University, Kaohiung 804, Taiwan}

\author{Bahadur~Singh}
\affiliation{Centre for Advanced 2D Materials and Graphene Research Centre, National University of Singapore, Singapore;}
 \affiliation{Department of Physics, National University of Singapore, 2 Science Drive 3, 117546, Singapore}

\author{Di~Wu}
\affiliation{Centre for Advanced 2D Materials and Graphene Research Centre, National University of Singapore, Singapore;}
 \affiliation{Department of Physics, National University of Singapore, 2 Science Drive 3, 117546, Singapore}

\author{Tay-Rong~Chang}
\affiliation{Department of Physics, National Cheng Kung University, Tainan 701, Taiwan}

\author{Titus~Neupert}
\affiliation{Department of Physics, University of Zurich, Winterthurerstrasse 190, 8057 Zurich, Switzerland}

\author{Su-Yang~Xu$^{\dag}$}
\affiliation{Laboratory for Topological Quantum Matter and Spectroscopy (B7), Department of Physics, Princeton University, Princeton, New Jersey 08544, USA}

\author{Hsin~Lin$^{\dag}$}
\affiliation{Centre for Advanced 2D Materials and Graphene Research Centre, National University of Singapore, Singapore;}
 \affiliation{Department of Physics, National University of Singapore, 2 Science Drive 3, 117546, Singapore}
\affiliation{Institute of Physics, Academia Sinica, Taipei 11529, Taiwan}

\author{M.~Zahid~Hasan$^{\dag}$\footnote[0]{$^{\dag}$Corresponding authors (emails): suyangxu@princeton.edu, nilnish@gmail.com, mzhasan@princeton.edu }}\affiliation{Laboratory for Topological Quantum Matter and Advanced Spectroscopy (B7), Department of Physics, Princeton University, Princeton, New Jersey 08544, USA}
\affiliation{Lawrence Berkeley National Laboratory, Berkeley, California 94720, USA}


\date{\today}

\maketitle

\textbf{Chiral crystals are materials whose lattice structure has a well-defined handedness due to the lack of inversion, mirror, or other roto-inversion symmetries~\cite{FlackH}.  These crystals represent a broad, important class of quantum materials; their structural chirality has been found to allow for a wide range of phenomena in condensed matter physics, including skyrmions in chiral magnets~\cite{Sky1}, unconventional pairing in chiral superconductors~\cite{chiral_sup1}, nonlocal transport and unique magnetoelectric effects in chiral metals~\cite{Nonlocal,Murakami_current, Morimoto},  as well as enantioselective photoresponse~\cite{Circular}. Nevertheless, while these phenomena have been intensely investigated, the topological electronic properties of chiral crystals remain largely uncharacterized.  While recent theoretical advances have shown that the presence of crystalline symmetries can protect novel band crossings in 2D and 3D systems \cite{GrapheneReview, ARPESrev, STMrev, 3DDirac1,Weyl2, Wan, Burkov2011, manes, Multi.Weyl,doubleWeyl2, doubleWeyl3, YoungkukLineNode, NS-3DDirac, WPVZ, DDP,QuantumChemistry,Sym_AV, NewF,RhSi,CoSi, QuantumChemistry, manes,NewF,RhSi,CoSi}, we present a new class of Weyl fermions enforced by the absence of particular crystal symmetries. These  ``Kramers-Weyl'' fermions are a universal topological electronic property of all nonmagnetic chiral crystals with spin-orbit coupling (SOC); they are guaranteed by lattice translation, structural chirality, and time-reversal symmetry, and unlike conventional Weyl fermions, appear at time-reversal-invariant momenta (TRIMs).   We cement this finding by identifying representative chiral materials in the majority of the 65 chiral space groups in which Kramers-Weyl fermions are relevant to low-energy physics.   By combining our analysis with the results of previous works~\cite{manes,DDP,QuantumChemistry,NewF,Multi.Weyl,doubleWeyl2,doubleWeyl3,RhSi,CoSi}, we determine that all point-like nodal degeneracies in nonmagnetic chiral crystals with relevant SOC carry nontrivial Chern numbers. We further show that, beyond the previous phenomena allowed by structural chirality~\cite{Sky1, chiral_sup1, Nonlocal,Murakami_current, Morimoto, Circular}, Kramers-Weyl fermions enable unusual phenomena, such as a monopole-like electron spin texture and chiral bulk Fermi surfaces over large energy windows.}

\begin{figure}[t]
\includegraphics[width=16cm]{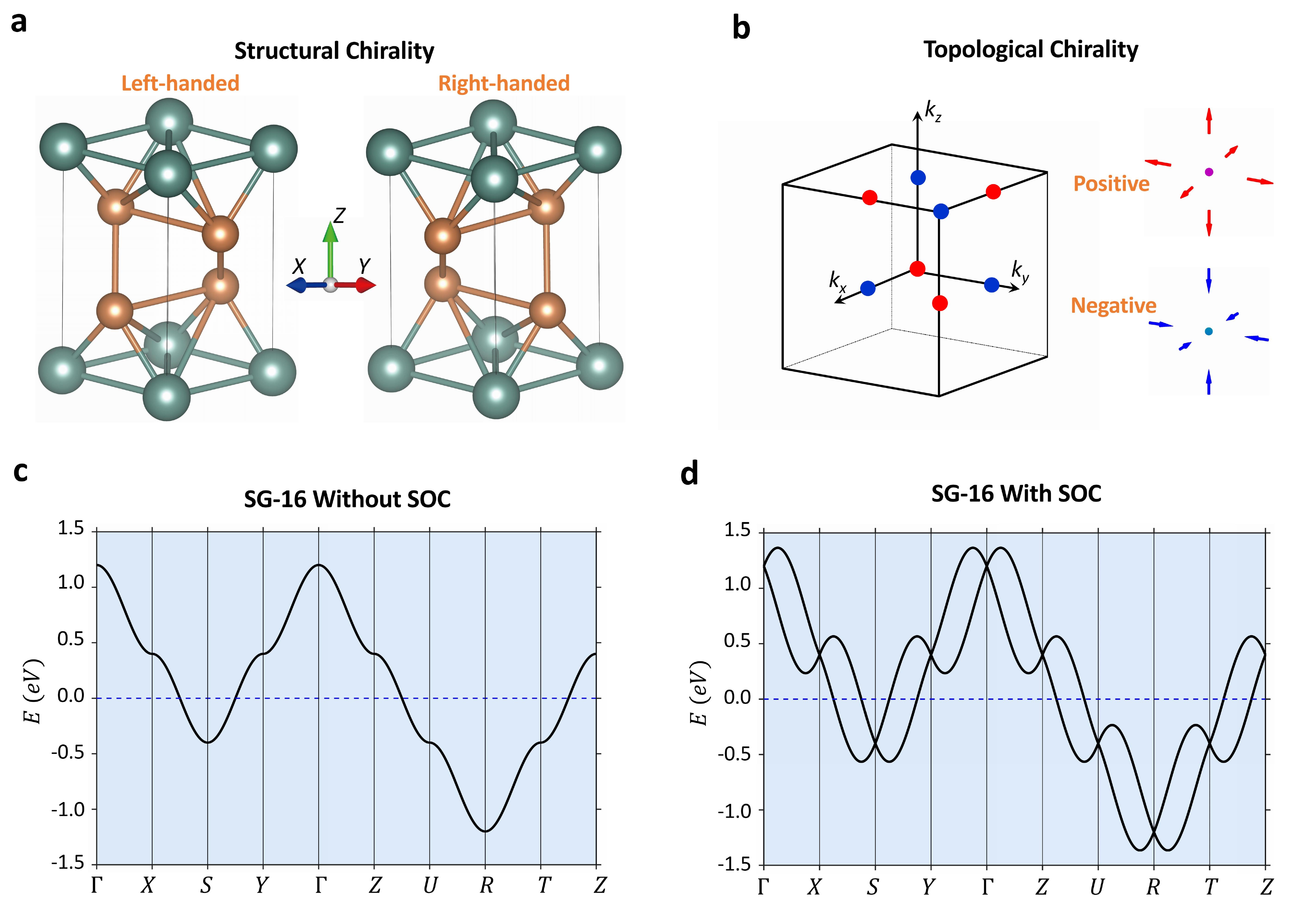}
\caption{{\bf Structural chirality and topological chirality.} ({\bf a}) Structurally chiral crystals have a distinct handedness, and are therefore characterized by an absence of inversion, mirror, or other roto-inversion symmetries~\cite{FlackH}. ({\bf b}) Topological chiral fermions act as monopoles or antimonopoles of Berry curvature. They are characterized by a quantized chiral charge, i.e., the quantized Chern number of the occupied bands on a closed $k$-surface surrounding the chiral fermion. In this paper, we find all nonmagnetic chiral crystals with SOC host topological chiral fermions at their TRIMs. ({\bf c}) The band structure for the SG-16 tight-binding model in the absence of SOC is characterized by a single band with a two-fold spin degeneracy. ({\bf d}) The inclusion of SOC splits the bands of this structurally chiral model everywhere except at the TRIMs, where Kramers theorem mandates that the bands remain doubly degenerate.  We find that these nodal degeneracies carry the same quantized chiral charge as conventional Weyl fermions, and therefore designate them ``Kramers-Weyl'' fermions.}
\label{Fig1}
\end{figure}

The spatial structures of three-dimensional crystal lattices are characterized by only a finite set of possible symmetries, giving rise to the 230 space groups (SGs) for nonmagnetic materials~\cite{BigBook}. In this work, we examine the properties of the SGs that characterize crystal structures with a sense of handedness, or structural chirality (Fig.~\ref{Fig1}\textbf{a}). Spatial inversion; mirror reflection; and roto-inversion, i.e., a combination of inversion and rotation, all invert structural chirality in crystals.  Of these 230 groups, there are 65 SGs free of chirality-inverting symmetries.  These  chiral SGs correspondingly characterize structurally chiral lattices~\cite{note}.

The electronic properties of structurally chiral crystals have been previously recognized as supporting a wide range of phenomena: chiral magnets support skymions~\cite{Sky1}, chiral metals show nonlocal and nonreciprocal electron transport~\cite{Nonlocal,Murakami_current}, chiral crystals further exhibit optical activity and magnetochiral dichroism~\cite{Circular}.   Because of these unusual properties, it is of great interest to search for possible topological electronic properties in chiral crystals.   A few previous studies (such as Ref.~\cite{Witczak2014}) have hinted at the existence of time-reversal invariant Weyl nodes in models that respect the symmetries of a chiral space group.  However, a systematic  understanding of the topological electronic properties of chiral crystals has remained elusive.   While  most of the recent theoretical advances are concerned with topological band structures related to the presence of additional crystalline symmetries (e.g., rotation, reflection, nonsymmorphic symmetries)~\cite{GrapheneReview, 3DDirac1,Weyl2, Wan, Burkov2011, manes, NS-3DDirac, WPVZ, DDP,QuantumChemistry,Sym_AV, NewF,RhSi,CoSi, QuantumChemistry, manes,NewF,RhSi,CoSi},  we highlight a class of nodal fermions enforced by structural chirality, and therefore by the absence of particular crystal symmetries.  In this work,  we show that structural chirality leads to a universal topological electronic property of all nonmagnetic chiral crystals with spin-orbit coupling (SOC): Kramers-Weyl fermions.

\color{black}
\bigskip
\textbf{I. Tight-Binding Model of a Spin-Orbit Coupled Chiral Crystal}

Without loss of generality, we present a tight-binding (TB) model in the symmorphic chiral SG 16 ($P222$) as a representative example of this physics.  Examining this model at the $\bs{k} \cdot \bs{p}$ level, we show that in SG 16, all Kramers degeneracies at TRIMs are Kramers-Weyl nodes with quantized chiral charge $|C|=1$.  We then use group theory in this section and enumerate irreducible representations in SM \textbf{B} to generalize this result, demonstrating that the same arguments apply to all of the TRIMs in symmorphic chiral SGs.

SG 16 characterizes a noncentrosymmetric orthorhombic crystal structure with two-fold rotation symmetries along each principle axis $x, y, z$. We consider a $1$-site unit cell on a primitive orthorhombic lattice with a spin-1/2 degree of freedom on each site. Considering all symmetry-allowed nearest neighbor hopping terms, the TB Hamiltonian reads
\begin{equation}
  \mathcal{H}(\bs{k}) = \displaystyle\sum_{i=x,y,z} t_{i}^{1}\cos(k_{i}) + t_{i}^{s}\sin(k_{i})\sigma^{i},
 \label{eq:SG16}
 \end{equation}
with $t_{i}^{1} \neq t_{j}^{1} \neq t_{i}^{s} \neq t_{j}^{s}$ $\forall$ $i \neq j$. Here $t_{i}^{1}$ denotes the $s$-orbital-like hopping strength and $t_{i}^{s}$ is a spin-orbit term. By examining the bands at each TRIM at the $\bs{k} \cdot \bs{p}$ level, we observe that each TRIM hosts a Weyl node described by
\begin{equation}
  \mathcal{H}_{\bs{k} \cdot \bs{p}} = \displaystyle
  \sum_{i=x,y,z} \left[u_{i}\left(1-\frac{k_{i}^2}{2}\right) + v_{i} k_{i}\sigma^{i}\right],
  \label{eq: kdotp}
\end{equation}
where in our model each TRIM has the same magnitude of $u_{i}$ and $v_{i}$ inherited from the lattice hopping parameters, but differing, TRIM-specific signs.  The wavefunction for the lower band, which we denote as $|{u}^{-}(\bs{k})\rangle$ for an arbitrary spin-1/2 system, exhibits a linear relationship between the trigonometric functions of the angles in spin space and $\bs{k}$. Hence, when integrated around a sphere $\mathcal{S}$ enclosing the Weyl node, the Berry curvature $\mathcal{F}$ of the lower band also wraps around the surface of the sphere exactly once. Therefore, the magnitude of the Chern numbers $C=\frac{1}{2\pi}\int{\mathcal{F}d\mathcal{S}}$ of the Kramers-Weyl nodes is $|C|=1$ at all TRIMs. In Figs.~\ref{Fig1}\textbf{c},\textbf{d}, we show the resulting band structure for this SG 16 TB model in the absence and presence of spin-orbit coupling (SOC), respectively.  As expected, the inclusion of SOC splits the two-fold-degenerate band in Fig.~\ref{Fig1}\textbf{c} and generates isolated two-fold-degenerate Kramers-Weyl nodes at all of the TRIMs (Fig.~\ref{Fig1}\textbf{d}) \cite{WPVZ}.  To determine the chiral charge of each Kramers-Weyl node, we examine the direction of the wrapping of $\mathcal{F}$ given by the relative signs of the \{$v_i$\} (the overall sign exchanges the Bloch states of the upper and lower bands, $|{u^{+}(\bs{k})}\rangle$ and $|{u^{-}(\bs{k})}\rangle$) to find the Chern number
\begin{equation}
C= \displaystyle\prod_{i=x,y,z}\text{sgn}(v_{i}).
 \end{equation}
This result indicates the chiral charge $C_{+1}=+1\times \text{sgn}(t_{x}^{s} t_{y}^{s} t_{z}^{s})$ for the Kramers pairs at TRIM points $\Gamma$, $S$, $U$, and $T$ and $C_{-1}=-1 \times \text{sgn}(t^s_x t^s_y t^s_z)$ for the pairs at  $X$, $Y$, $Z$, and $R$, as illustrated in Fig.~\ref{Fig1}\textbf{b}. The sign of $C$ at $\Gamma$ reflects the signs of the \{$t^s_i$\} for the position space lattice. Thus, the chirality of the atomic positions is directly responsible for defining the handedness of the Kramers-Weyl node at $\Gamma$ and therefore, by the pattern of alternating signs, at all of the other TRIMs as well.

As our determination of $C=\pm1$ only relied on the two-fold degeneracy and linear dispersion of Eq.~\eqref{eq: kdotp},  we deduce that any such degeneracy must be a Weyl node.  More precisely, as the little groups of all eight TRIMs in SG 16 are isomorphic to the same chiral point group: $222$, which with spinful $\mathcal{T}$ symmetry only has a single two-dimensional corepresentation~\cite{BigBook, QuantumChemistry}, we conclude that any TRIM with a little group isomorphic to point group $222$ must also host Kramer-Weyl fermions.  Tuning parameters such that $u_{i}=0,\ v_{i}=v_{j}$ for all $i$,$j$, this $\bs{k} \cdot \bs{p}$ theory becomes isotropic, and thus invariant under the action of any chiral point group.  Taken together, this implies that the little group of the $\Gamma$ point in an arbitrary chiral crystal, which is isomorphic to the point group of that crystal, and every TRIM point in symmorphic chiral crystals, which host symmetry algebras unmodified by the projections of fractional lattice translations, must always allow at least one irreducible corepresentation of a Kramers-Weyl fermion.  In SM \textbf{B}, we explicitly confirm this conclusion and list in the language of Refs.~\cite{BigBook,NewF,DDP,RhSi} the irreducible corepresentations that describe Kramers-Weyl fermions.  We also show in SM \textbf{C},\textbf{D} that Kramers-Weyl fermions with $|C|=3$ and nonlinear dispersion are permitted in chiral point groups with three- and six-fold rotation symmetries, and in SM \textbf{D} identify the irreducible corepresentations of these higher-Chern-number Kramers-Weyl fermions.  Previous works have also exploited similar simple models in discussing Weyl fermions~\cite{Witczak2014,BernevigTalk}.  However, those works did not explore the relations of the little groups at the TRIMs with each other, and more importantly did not recognize that their isomorphisms to chiral point groups allow for generalizing to other crystal systems.

In nonsymmorphic chiral crystals, these arguments become modified away from the $\Gamma$ point, and the Kramers-Weyl nodes can become obscured by additional band degeneracies.  Consider, for example, the chiral SG 19 ($P2_{1}2_{1}2_{1}$), which differs from SG 16 in that all of its two-fold rotations are instead nonsymmorphic screws: $s_{x}=\{\mathcal{C}_{2x}|\frac{1}{2}\frac{1}{2}0\}$, $s_{y}=\{\mathcal{C}_{2y}|0\frac{1}{2}\frac{1}{2}\}$, and $s_{z}=\{\mathcal{C}_{2z}|\frac{1}{2}0\frac{1}{2}\}$. In reciprocal space, $s_{i}\times \mathcal{T}$ are valid symmetry operations that map a $\bs{k}$ point to itself on the $k_{i}=0, \pi$ planes.  When acting on a Bloch eigenstate of the Hamiltonian, $(s_{i}\mathcal{T})^{2}=e^{-i k_i}$ enforces two-fold Kramers degeneracies throughout all three $k_{i}=\pi$ planes. TRIMs belonging to exactly two nodal planes will host four-fold degeneracies, due to the algebra of the screw rotations at those $\bs{k}$ points. The modified algebra of the screw rotations at $k_{i}=k_{j}=\pi$ can also be understood by noting that the little groups of those TRIMs are no longer isomorphic to those at $\Gamma$, as they were in SG 16, due to the projective effects of the fractional lattice translations on the commutation relations of the symmetry representations~\cite{NS-3DDirac}. In SM \textbf{E},\textbf{F}, we more closely study these nodal surfaces, using the treatment prescribed in Ref.~\cite{NodalSurfaces} to determine that they have a nonzero chiral charge.  We explore these chiral nodal surfaces with an explicit TB model for SG 19, in analogy to the model presented in this section for SG 16, finding that in SG 19, the nodal planes must necessarily carry an odd Chern number.

We also note that in certain achiral SGs, it is possible for some of the TRIMs to still have little groups isomorphic to chiral point groups due to the relationship between the SG roto-inversion axes, mirror planes, and the reciprocal lattice vectors.  In SM \textbf{G}, we explore this possibility and show that Kramers-Weyl fermions still occur in achiral crystals of BiTeI in SG 156 (Inorganic Crystal Structure Database (ICSD) \#10500~\cite{ICSD}).

It is important to highlight that in the context of symmetry and topology, Kramers-Weyl fermions are distinct from previous examples of chiral fermions.  In conventional band-inversion Weyl semimetals \cite{Weyl2, Wan, Burkov2011}, such as TaAs~\cite{ARPES-TaAs1, ARPES-TaAs2, ARPES-TaAs3}, Weyl nodes need not be protected by symmetry~\cite{QuantumChemistry}, and instead maybe pairwise nucleated or annihilated without breaking time-reversal ($\mathcal{T}$) or crystal symmetries by removing the band inversion.  The presence of Weyl nodes in band-inversion semimetals is thus not guaranteed by general symmetry criteria, but is instead the result of favorable energetics.  More exotic chiral fermions also comprise a subset of the recently proposed ``unconventional fermions''~\cite{manes,NewF, DDP, RhSi,CoSi}, which are guaranteed to exist in the electronic structures of certain crystals under highly specific symmetry and orbital conditions~\cite{QuantumChemistry}.  However, due to this specificity, these unconventional higher-degeneracy fermions are not robust to perturbations that break the exact symmetries of their SGs~\cite{DDP, Triple1,Triple2, Triple3}, and very few experimentally viable material candidates have been presented to date, with the RhSi family standing out as a notable exception~\cite{RhSi, CoSi}.  Kramers-Weyl fermions, conversely, are neither generated by a band inversion, nor are dependent on specific combinations of nonsymmorphic crystal symmetries~\cite{manes,NewF, DDP,RhSi}; they are guaranteed to exist, and to be chiral fermions, merely by the action of $\mathcal{T}$ symmetry on the irreducible representations of the chiral point groups.  Moreover, as $|C|=1$ Weyl fermions, they can only be destroyed through pairwise annihilation, which may only occur under a large magnetic field that moves them far off from the TRIMs, or through the zone-folding effects of commensurate antiferromagnetism or charge-density waves.    Finally, Ref. \cite{QuantumChemistry} showed that, apart from conventional $|C|=1$ Weyl  fermions, all degeneracies in nonmagnetic crystals are captured by the set of irreducible (co)represenations.  For each of these degeneracies, a separate calculation of the chiral charge can then be performed.  Remarkably, by exhaustively enumerating all of the irreducible (co)represenations of the 65 chiral SGs, and comparing them to the set of known chiral fermions, now including the Kramers-Weyl fermions highlighted in this work, we find that all point degeneracies in chiral SGs exhibit nonzero chiral charge  (SM~{\bf N}).  Specifically, they are either conventional $|C|=1$ Weyl fermions ~\cite{Burkov2011,Wan,Weyl2}, double- or triple-Weyl fermions on rotation axes~\cite{Multi.Weyl, doubleWeyl2,doubleWeyl3}, unconventional chiral fermions at high-symmetry crystal momenta~\cite{NewF,RhSi,CoSi}, or Kramers-Weyl fermions.

From a theoretical perspective, intrinsic-filling Kramers-Weyl metals also represent the simplest examples of filling-enforced semimetals~\cite{WPVZ}.  Though previous works have focused on the role of nonsymmorphic symmetries in generating large and exotic band connectivities~\cite{WPVZ,NewF,RhSi,QuantumChemistry}, we find in this work that even the simplest filling restriction, that a $\mathcal{T}$-symmetric crystal with an odd number of electrons per unit cell must be gapless, can be used to predict materials with Weyl fermions.  We note that some of the chiral materials examined in this work are narrow band gap semiconductors. In those cases, despite an intrinsic insulating filling, a moderate chemical doping can shift the chemical potential into the conduction or valence bands and isolate a Kramers-Weyl node (see Fig.~\ref{Fig3} and relevant discussions below).  A deeper theoretical understanding of this preference may provide greater insight into the appearance of Mott insulating and other interacting phases in certain filling-enforced semimetals~\cite{CopperMott,InteractingDDP}. \color{black}

\bigskip
\textbf{II. Kramers-Weyl Physics in Previously Synthesized Chiral Materials}

Consulting the ICSD~\cite{ICSD}, we identify in Tables~\ref{symChiral} and~\ref{nonsymChiral} representative examples of previously synthesized chiral crystals and highlight in blue the materials in which Kramers-Weyl fermions are relevant to low-energy physics, which we demonstrate with calculated electronic structures provided in SM \textbf{J}.  The number of readily synthesizable materials varies greatly across the 65 chiral space groups.  We find that SGs 4, 19, 173, and 198 characterize large numbers of known chiral crystals, whereas SGs 89, 93, 153, 171, and 172 are devoid of promising material candidates.
As no systematic experimental exploration of Weyl fermions in chiral crystals has been performed yet, these tables should provide a guide towards future transport and spectroscopic analyses.

Among the chiral materials listed in Tables~\ref{symChiral} and~\ref{nonsymChiral}, there are both chiral metals, as well as insulators.  Though the Kramers-Weyl nodes in insulators are not generically relevant to transport experiments, they may still be probed with angle-resolved photoemission spectroscopy (ARPES)~\cite{ARPES-TaAs1, ARPES-TaAs2, ARPES-TaAs3}, scanning tunneling microscopy \cite{Cd3As2_Ali}, resonant inelastic x-ray scattering \cite{RIXS}, neutron scattering \cite{Nagaosa_neutron} or by other optical measurements \cite{Q_photo_Cur}.  Furthermore, when the Kramers-Weyl nodes of an insulator are near the Fermi level, they could also be studied in transport experiments \cite{Ong} after slight electron or hole doping, which we discuss further below in the context of Fermi arc observation and in Sec.~III with regards to the circular photogalvanic response effect (CPGE).

We present two material candidates particularly representative of this physics: a chiral metal and a chiral insulator with spectroscopically accessible Kramers-Weyl Fermi arcs.  The band structure of the symmorphic compound Ag$_3$BO$_3$(\#26521) (SG 155) is plotted in Figs.~\ref{Fig2}\textbf{b} and~\textbf{c} without and with SOC, respectively; we highlight the Kramers-Weyl nodes with orange circles. In SM \textbf{E}, we explore in more detail chiral metallic phases in nonsymmorphic chiral crystals, highlighting in particular Rh(Ir)Ge(Sn)$_{4}$ in SG 152.  As in conventional Weyl semimetals, Kramers-Weyl points also give rise to Fermi arcs which, on surfaces for which bulk TRIMs project onto surface TRIM points, must necessarily appear in time-reversed pairs. We identify the insulating compound AgBi(Cr$_2$O$_7$)$_2$ (\#14233) in SG 79 as representative of this physics.  The bulk electronic structure of AgBi(Cr$_2$O$_7$)$_2$ in the absence and presence of SOC is shown in Figs.~\ref{Fig2}\textbf{f} and~\textbf{g}, respectively. We confirm the existence of time-reversed pairs of Fermi arcs $\sim -0.23 $ eV below the Fermi energy by performing a slab calculation for a surface termination along the (110) crystallographic direction, shown in Fig.~\ref{Fig2}\textbf{i}.  As shown in Fig.~\ref{Fig2}\textbf{j}, a loop taken around one of the surface TRIMs exhibits a projected Chern number of $C=+2$, necessitating the existence of two Fermi arcs. These Fermi arcs can be directly observed by ARPES, as they lie below the Fermi level.  We further note that, because the Kramers-Weyl nodes are pinned to the TRIMs, chiral crystals can in principle support the longest possible Fermi arcs, and therefore span the entire surface BZ like those in the unconventional chiral semimetal  RhSi~\cite{RhSi,CoSi}.  However, this is in fact very challenging to realize in real materials, for which the band widths, determined by the hopping amplitudes, are typically much greater than the spin splitting, which is determined by the strength of SOC (SM \textbf{I}).

\textbf{III. Novel Phenomena in Chiral Crystals}

 Below we describe five novel phenomena relevant to the Kramers-Weyl fermions in chiral crystals. Phenomena \#1 and \#2 are unique to Kramers-Weyl fermions and have not been previously proposed. Phenomena \#3--\#5 have been proposed by previous works and are not unique to Kramers-Weyl fermions. Instead, we highlight how Kramers-Weyl fermions provide a new and previously unrecognized venue for realizing these phenomena.

\emph{1. Spin texture of Kramers-Weyl fermion}

The most general Hamiltonian of a $|C|=1$ Weyl fermion can be written as $\mathcal{H}_{\text{Weyl}}(\bs{k}) = v_ik_i\sigma_0+ A_{ij} k_{i} \sigma^{j}$, where $A_{ij}$ and $v_i$ are real numbers and sums over repeated indices $i,j=x,y,z$ are implied.
For the Weyl nodes in a band-inversion Weyl semimetal, $\bs{\sigma}$ represents only an effective pseudo-spin degree of freedom, which is very difficult to measure in a momentum resolved fashion. The physical spin, which can be directly measured by spin-resolved ARPES, is distinct from this pseudo-spin.

Conversely, in the $\bs{k} \cdot \bs{p}$ theory of Kramers-Weyl fermions, $\bs{\sigma}$ represents the true electron spin in the limit in which the energy scale of SOC is much smaller than the interband separation in absence of SOC (see SM \textbf{L}). Remarkably, we find that, in this limit, the chiral charge of a Kramers-Weyl fermion can be directly probed by measuring the spin texture. This is a unique property of Kramers-Weyl fermions. Specifically, the physical spin on Fermi surfaces enclosing a Kramers-Weyl node sweeps out the full unit sphere. Since the chiral charge is given by $C=\mathrm{sgn}\,[\mathrm{det}\,(A)]$, $C$ can be directly obtained by measuring the spin polarization $\bs{S}_{\bs{k}}=\langle \bs{k}|\bs{\sigma}|\bs{k}\rangle$ near the Kramers-Weyl nodes by calculating
\begin{equation}
	C^{\textrm{Kramers-Weyl}}=-\mathrm{sign}(k)\mathrm{sign}(\bs{S}_{(k,0,0)}\times \bs{S}_{(0,k,0)}\cdot \bs{S}_{(0,0,k)}).
\end{equation}
In addition, $v_i=0$ for Kramers-Weyl fermions by time-reversal symmetry, so that the Kramers-Weyl-cone cannot be `tilted' in momentum space.
Moreover, in the presence of additional rotational symmetries, Kramers-Weyl fermions exhibit spin-momentum locking (see SM \textbf{L} for a detailed discussion). Specifically, for a $C=+1$ Kramers-Weyl node, the spin either points outward along all three principal directions (e.g., $k_x$, $k_y$, $k_z$) or it points outward along one direction and inward along the other two directions; for a $C=-1$ Kramers-Weyl node, the spin either points inward along all three principal directions, or it points inward along one direction and outward along the other two directions.  In addition, $\mathcal{T}$ enforces $v_i=0$ for Kramers-Weyl fermions, thus the Kramers-Weyl cone cannot be `tilted' in momentum space \cite{typeii}.

Using first-principles calculations, we confirm the presence of this spin texture  in the Kramers-Weyl fermions in Ag$_2$Se, and contrast it with that of the conventional band-inversion induced Weyl fermions in TaAs (Figs.~\ref{Fig3}\textbf{e}, \textbf{f}). In SM \textbf{M}, we further show that the real spin-momentum locking of Kramers-Weyl nodes can lead to a large spin Hall conductivity.

\color{black}
\emph{2. Chiral bulk Fermi surfaces over large energy windows}

A Fermi pocket that encloses a single Weyl node carries a quantized Chern number. Recent works~\cite {Chris, Berry_phase, CDW} have highlighted that such Fermi pockets can lead to unique transport and symmetry-breaking phenomena, including novel magnetic breakdowns \cite{Chris}, unconventional quantum oscillations \cite{Berry_phase}, giant spin Hall effects \cite{spin_Hall}, and chiral charge density waves \cite{CDW}.

For a conventional band-inversion induced Weyl semimetal, isolated chiral Fermi surfaces may only be realized at energies close to the Weyl nodes, which typically are only separated by a few hundredths of the BZ in momentum, and by of the order or less than 100meV in energy~\cite{ARPES-TaAs1, ARPES-TaAs2, ARPES-TaAs3}.  Therefore it is quite challenging to realize band-inversion Weyl semimetals with isolated chiral Fermi surfaces; only TaAs and TaP display well-isolated chiral surfaces, whereas in other materials, such as NbP, NbAs, WTe$_2$ and MoTe$_2$, the chemical potential misses the narrow energy window defined by the weak band inversion.

In stark contrast, in chiral crystals, isolated Fermi surfaces with nonvanishing Chern number can form at any energy between the highest and the lowest Kramers-Weyl nodes or unconventional chiral fermions in a set of connected bands. The scale of this energy window is governed by the lattice hopping, typically much larger than the accessible scale of band inversion in unstrained crystals like conventional Weyl semimetals, and the splitting between Fermi surfaces of opposite Chern number is instead determined by the strength of SOC (Fig.~\ref{Fig1}\textbf{d} and SM \textbf{H},\textbf{I}).

\emph{3. Quantized circular photogalvanic current}

In a recent theoretical work, de Juan \textit{et al.} proposed a quantized photogalvanic response in roto-inversion-free Weyl semimetals~\cite{Q_photo_Cur}, where Weyl nodes of one chirality sit at the Fermi level, and nodes of the opposite chirality lie energetically far away from it. A crucial limitation that has hindered the realization of this phenomena has been the lack of viable materials platforms identified to date~\cite{Q_photo_Cur}. Indeed, very few Weyl semimetals are known to satisfy this stringent requirement.
 In contrast, Kramers-Weyl fermions  with opposite chirality are not related by crystal symmetries, and absent fine tuning will appear in real materials at different energies.  \color{black} Thus, the universal presence of Kramers-Weyl fermions in chiral crystals allows access to a simple, powerful strategy for identifying additional ideal material candidates for the observation of quantized photocurrent: identify semimetallic chiral crystals where a Kramers-Weyl node is isolated at the intrinsic chemical potential, or identify narrow-gap semiconducting chiral crystals where a Kramers-Weyl node can be isolated by moderate electron or hole doping.

\color{black}
We present Ag$_2$Se$_{0.3}$Te$_{0.7}$ (SG 19) as an example for doping-enabled photocurrent (Fig.~\ref{Fig3}). As discussed in Sec.~I, crystals in SG 19 display chiral Kramers-Weyl nodes at $\Gamma$, and two-fold-degenerate chiral nodal planes along the BZ boundaries ($k_i=\pi$) (Figs.~\ref{Fig3}\textbf{a,b}).  With slight electron doping, one could shift the Fermi level to the Kramers-Weyl fermions (the blue dashed line). The calculated photogalvanic current of Ag$_2$Se$_{0.3-\delta}$Te$_{0.7}$ ($\delta$=0.00016) (Fig.~\ref{Fig3}\textbf{d}) exhibits a quantized value in the terahertz (THz) photon energies ($\sim1$ meV), which are experimentally accessible~\cite{Heinz}.  Continuing with this strategy, we easily identify a large number of additional chiral materials platforms for the observation of Kramers-Weyl-enabled quantized photocurrent: CsCuBr$_{3}$(\#10184) in SG-20, BaAg$_2$SnSe$_4$(\#170856) in SG-23, NbO$_2$ (\#35181) in SG-80, Ag$_3$SbO$_4$(\#417675) in SG-91, MgAs4(\#1079) in SG-92, Cu$_{2}$S(\#16550) in SG-96, Ta$_{2}$Se$_{8}$I(\#35190) in SG-97, CdAs$_{2}$ (\#16037) in SG-98, Ag$_3$IS(\#93431) in SG-146, TlTe$_{2}$O$_{6}$(\#4321) in SG-150, SrIr$_{2}$P$_{2}$(\#73531) in SG-154, BiO$_{2}$(\#27152) in SG-197, K$_{2}$Sn$_{2}$O$_{3}$(\#40463) in SG-199, and Ag$_{3}$Se$_{2}$Au(\#171959) in SG-214 are similar chiral crystals with narrow band gaps such that moderate doping may allow the isolation of a Kramers-Weyl node at the Fermi level; and BaCu$_2$Te$_2$O$_6$Cl$_2$ (\#85786) in SG-4 and Ca$_2$B$_5$Os$_3$(\#59229) in SG-5 are semimetallic crystals which each have Fermi pockets that enclose a single Kramers-Weyl node at intrinsic doping.  The band structures of these compounds are shown in SM \textbf{J}.

\emph{4. The chiral magnetic effects}

The chiral and gyrotropic effects, previously proposed in Refs.~\cite{gyrotropic,gyrotropic2}, are linear responses in Weyl semimetals to electromagnetic waves.  In these chiral magnetic effects, a dissipationless current arises in response to an alternating magnetic field.  These effects are crucially reliant on the same energetic restrictions as the CPGE in the previous section: that Weyl nodes of different chiralities lie at different energies.  Like the CPGE, the realization of these effects has been hindered due to the absence of ideal material platforms.  Our proposal of Kramers-Weyl fermions immediately provides ample new feasible platforms (all materials highlighted above) in which to probe these response effects.

\emph{5. Other exotic phenomena in chiral crystals}

Finally, we propose that Kramers-Weyl fermions may offer a new way to systematically control and modulate a number of novel symmetry-allowed phenomena in structurally chiral crystals, including the magnetochiral and magnetoelectric effects~\cite{Nonlocal,Murakami_current, Morimoto}. While these phenomena can still occur in a chiral crystal in the absence of a Weyl node near the Fermi energy, recent theoretical works have highlighted that the presence of strong Berry curvature can dramatically enhance the magnitude of these effects~\cite{Murakami_current, Morimoto}. In addition, the presence of Kramers-Weyl fermions near the Fermi energy may offer the possibility of studying superconducting pairing on Fermi surfaces with nonzero Chern number \cite{WSC, WSC2}, which may be a promising recipe for engineering unconventional superconductivity.

\bigskip

\color{black}
\bigskip
{\bf Acknowledgements}

Work at Princeton was supported by the US Department of Energy under Basic Energy Sciences (Grant No. DOE/BES DE-FG-02-05ER46200). M.Z.H. acknowledges Visiting Scientist support from Lawrence Berkeley National Laboratory, and partial support for theoretical work from the Gordon and Betty Moore Foundation (Grant No. GBMF4547/Hasan). The work at the National University of Singapore was supported by the National Research Foundation, Prime Minister's Office, Singapore under its NRF fellowship (NRF Award No. NRF-NRFF2013-03). B.J.W. acknowledges support through a Simons Investigator grant from the Simons Foundation to Charles L. Kane, through Nordita under ERC DM 321031, through grants from the Department of Energy (No. DE-SC0016239), the Simons Foundation (Simons Investigator Grant No. ONR-N00014-14-1-0330), the Packard Foundation, and the Schmidt Fund to B. Andrei Bernevig, and acknowledges the hospitality of the Donostia International Physics Center.  F.S. and T.N. acknowledge support by the Swiss National Science Foundation (grant number 200021-169061) and the ERC-StG-Neupert-757867-PARATOP, respectively. T.-R.C. was supported by the Ministry of Science and Technology under MOST Young Scholar Fellowship: MOST Grant for the Columbus Program NO. 107-2636-M-006-004-, National Cheng  Kung University, Taiwan, and National Center for Theoretical Sciences (NCTS), Taiwan. M.Z.H. acknowledges support from the Miller Institute of Basic Research in Science at the University of California at Berkeley in the form of Visiting Miller Professorship during the early stages of this work. The authors thank Charles L. Kane and Randall Kamien for helpful discussions on chirality and thank Barry Bradlyn, Jennifer Cano, Mois I. Aroyo, and B. Andrei Bernevig for insightful discussions on group theory and symmetry.


\newpage

\newpage
\begin{figure}[t]
\includegraphics[width=18cm]{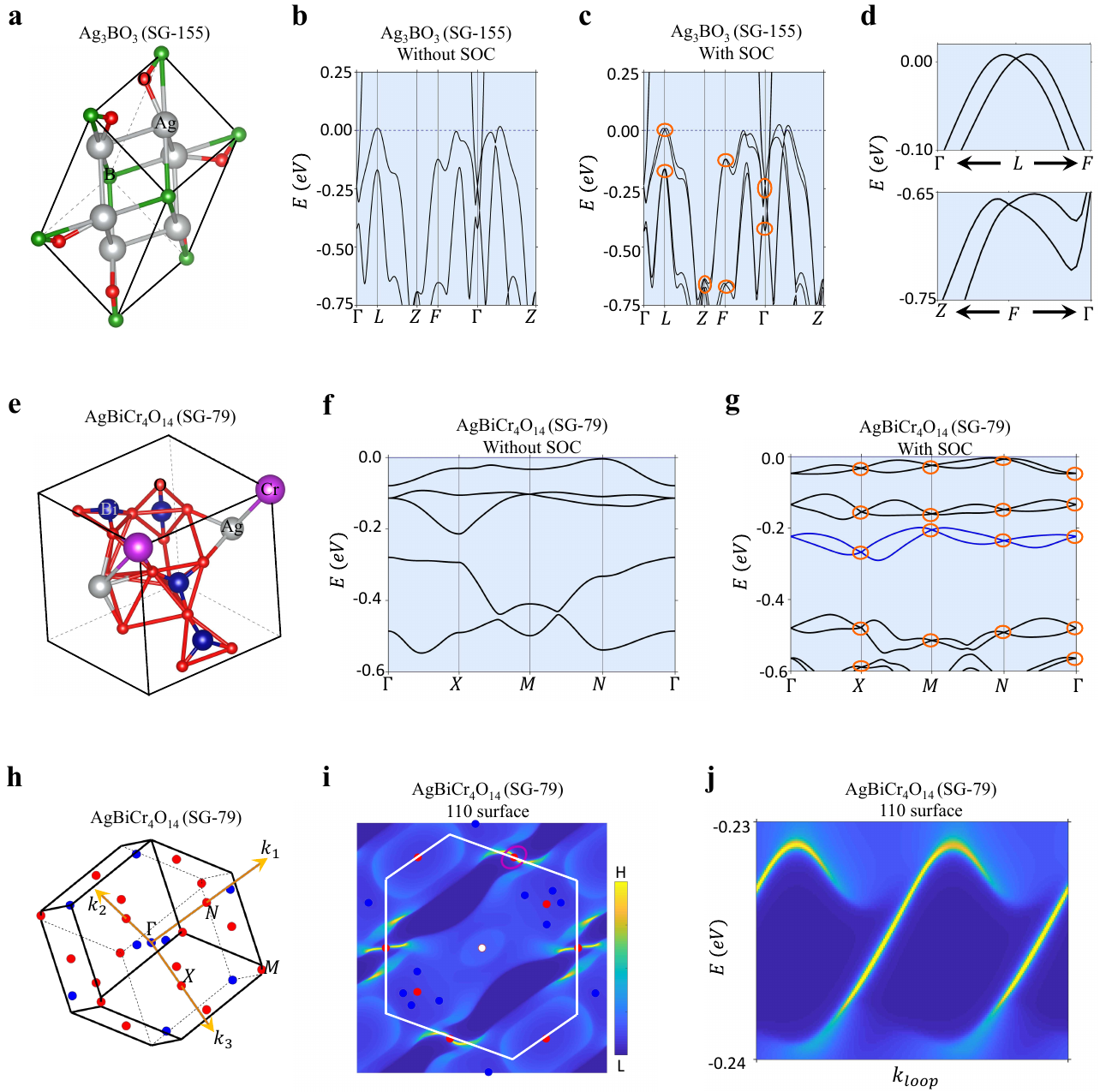}
\caption{{\bf Electronic structure and Fermi arcs of Kramers-Weyl material candidates.}
({\bf a}) The chiral crystal Ag$_3$BO$_3$ in SG 155.
({\bf b, c}) The band structures of Ag$_3$BO$_3$ passing through TRIMs in the absence and in presence of SOC, respectively. All of the TRIM points of Ag$_3$BO$_3$ host isolated Kramers-Weyl nodes in the presence of SOC, as indicated by the orange circles.
({\bf d}) The Kramers-Weyl band structures in the vicinity of $F$ and $L$. ({\bf e-g}) The crystal, electronic, and zoomed-in band structures of AgBi(Cr$_2$O$_7$)$_2$ in SG-79.}
\label{Fig2}
\end{figure}
\addtocounter{figure}{-1}
\begin{figure*}[t!]
\caption{({\bf h}) Momentum space distribution and chiralities of the Weyl nodes in AgBi(Cr$_2$O$_7$)$_2$ between the two blue bands. The red and blue circles indicate Weyl nodes of opposite chiral charges. ({\bf i})  The (110) surface state spectral function of AgBi(Cr$_2$O$_7$)$_2$ at $E=-0.235$~eV. The Fermi arc surface states connecting the projections of the two distinct bulk pockets clearly visible. The projected charges of red (blue) circles are $+2$ ($-1$). ({\bf j})  The surface spectral function along the $k$-path highlighted in panel (i) exhibits a $+2$ winding number, consistent with the pair of chiral Fermi arcs observed in (i).}
\label{Fig2}
\end{figure*}

\begin{center}
\begin{table}[t]

\begin{tabular}{|>{\columncolor{mygray}\sf}m{1.2cm}<{\centering}|>{\columncolor{mypink}\sf}c|>{\columncolor{mycay}\sf}m{2.2cm}<{\centering}|p{0.2cm}|>{\columncolor{mygray}\sf}m{1.2cm}<{\centering}|>{\columncolor{mypink}\sf}c|>{\columncolor{mycay}\sf}m{2.2cm}<{\centering}|}
\hline

\bf{Space Group} & \bf{Material} & \bf{ Number of Collections in the ICSD } & & \bf{Space Group} & \bf{Material} & \bf{ Number of Collections in the ICSD } \\
\toprule
\toprule
1 &  \blue{ Li$_{6}$CuB$_{4}$O$_{10}$(\#249215)} & 492 & & 146 &  \blue{$\beta$-Ag$_{3}$IS(\#93431)} & 307 \\
3 &  Pb$_{3}$GeO$_{5}$(\#200517) & 58 & & 149 &  RbGeIO$_{6}$(\#73613) & 40\\
5 & \blue{Ca$_{2}$B$_{5}$Os$_{3}$(\#59229)}  & 539 &  & 150 & \blue{Tl$_{2}$TeO$_{6}$(\#4321)} & 314\\
16 & AlPS$_{4}$(\#15910) & 13 & & 155 & \blue{Ag$_{3}$BO$_{3}$(\#26521)} & 244\\
21 & YSb$_{2}$(\#651733) & 51 &  & 168 & K$_{2}$Ta$_{4}$F$_{4}$O$_{9}$(\#8204) & 3\\
22 & \blue{ ThOs$_{2}$B$_{2}$(\#601346)} & 55 &  & 177 & PbS$_{2}$O$_{6}$ $\cdot$ 4H$_{2}$O(\#68630) & 3\\
23 & \blue{BaAg$_{2}$SnSe$_{4}$(\#170856)}  & 28 & & \multicolumn{1}{>{\columncolor{magenta}}c}{195} & SnI$_{4}$(\#18010) & 29\\
75 & \blue{K$_{4}$CuV$_{5}$ClO$_{15}$(\#401042)}  & 27 &  & \multicolumn{1}{>{\columncolor{magenta}}c}{196} & $\alpha$-Cu$_{2}$Se(\#59955) & 47\\
79 & \blue{AgBi(Cr$_{2}$O$_{7}$)$_{2}$(\#14233)} & 45 & & \multicolumn{1}{>{\columncolor{magenta}}c}{197} & \blue{m-Bi$_{2}$O$_{3} $(\#27152)} & 174\\
89 & - & 3 & & \multicolumn{1}{>{\columncolor{magenta}}c}{207} & RbNO$_{3}$(\#60966) & 1 \\
97 & \blue{Ta$_{2}$Se$_{8}$I(\#35190)} & 9 & & \multicolumn{1}{>{\columncolor{magenta}}c}{209} & Na$_{3}$PO$_{4}$(\#14090) & 8  \\
143 & RbW$_{3}$O$_{9}$(\#96421) & 112 & & \multicolumn{1}{>{\columncolor{magenta}}c}{211} & NiHg$_{4}$(\#151197) & 3  \\

\hline
\end{tabular}
\caption{ \textbf{Representative topological chiral crystals with Kramers-Weyl fermions in symmorphic chiral space groups.} We list all 24 symmorphic chiral space groups and, where possible, material candidates from the inorganic crystal structure database (ICSD)~\cite{ICSD}.  All the compounds listed in the table host Kramers-Weyl fermions. Among them, the compounds labeled in blue are (semi-)metals or small-gap insulators with clean Fermi surfaces and Kramers-Weyl fermions near the Fermi level; their electronic structures are shown in Supplemental Infromation \textbf{J}. The numbers in brackets are the ICSD collection codes. The space groups labeled in magenta may also host additional unconventional four-fold-degenerate chiral fermions at time-reversal-invariant momenta (TRIMs) with little groups isomorphic to chiral point groups.  Space groups 195, 196, and 197 have TRIM-point little groups isomorphic to chiral point group $23$ ($T$), and may thus host the unconventional chiral fermion identified in Refs.~\cite{RhSi,CoSi}.  Space groups 207, 209, and 211 have TRIM-point little groups isomorphic to chiral point group $432$ ($O$), and may thus host the unconventional spin-3/2 chiral fermion identified in Ref.~\cite{NewF}.}
\label{symChiral}
\end{table}
\end{center}

\begin{center}
\begin{table}

\begin{tabular}{|>{\columncolor{mygray}\sf}m{1.2cm}<{\centering}|>{\columncolor{mypink}\sf}c|>{\columncolor{mycay}\sf}m{2.2cm}<{\centering}|p{0.2cm}|>{\columncolor{mygray}\sf}m{1.2cm}<{\centering}|>{\columncolor{mypink}\sf}c|>{\columncolor{mycay}\sf}m{2.2cm}<{\centering}|}
\hline

\bf{Space Group} & \bf{Material} & \bf{ Number of Collections in the ICSD } & & \bf{Space Group} & \bf{Material} & \bf{ Number of Collections in the ICSD } \\
\toprule
\toprule
4 & \blue{ BaCu$_{2}$Te$_{2}$O$_{6}$Cl$_{2}$(\#85786)} & 912 & & 152 &  \blue{IrGe$_{4}$(\#53655)} & 408 \\
17 &  Ba$_{2}$Cu$_{3}$YPb$_{2}$O$_{8}$(\#66088) & 29 & & 153 &  - & 1 \\
18 & \blue{Pd$_{7}$Se$_{4}$(\#77897)} & 177  &  & 154 & \blue{SrIr$_{2}$P$_{2}$(\#73531)} & 159\\
19 & \blue{$\alpha$-Ag$_{2}$Se(\#261822)} & 1145 & & 169 & $\alpha$-In$_{2}$Se$_{3}$(\#82203) & 55\\
20 & \blue{CsCuBr$_{3}$(\#10184)}  & 219  & & 170 & BaN$_{2}$O$_{4}$ $\cdot$ H$_{2}$O(\#201484) & 14\\
24 & K$_{2}$PdSe$_{10}$(\#71947) & 10   & &  171 & - & 4\\
76 & \blue{TlBO$_{2}$(\#36404)} & 64 &  & 172 & - & 1\\
77 & MgB$_{2}$O(OH)$_{6}$(\#24920) & 9 & & 173 & \blue{CuLa$_{4}$S$_{7}$(\#628240)} & 1238\\
78 & Sr$_{2}$As$_{2}$O$_{7}$(\#190008) & 22 & &  178 & Hf$_{5}$Ir$_{3}$(\#638575) & 55\\
80 & \blue{$\beta$-NbO$_{2}$(\#35181)} & 22 & & 179 & Na$_{3}$B$_{4}$O$_{7}$Br(\#252106) & 32\\
90 & Na$_{4}$Ti$_{2}$Si$_{8}$O$_{22}$ $\cdot$ 4H$_{2}$O(\#240912) & 20 & & 180 & \blue{NbGe$_{2}$(\#16503)} & 241\\
91 & \blue{Ag$_{3}$SbO$_{4}$(\#417675)} & 31 & &  181 & \blue{WAl$_{2}$(\#173662)} & 48\\
92 & \blue{MgAs$_{4}$(\#1079)} & 309 & & 182 & \blue{PbRbIO$_{6}$(\#73615)} & 217\\
93 & - & - & &    \multicolumn{1}{>{\columncolor{magenta}}c}{198} & \blue{$\beta$-RhSi(\#79233)} & 766\\
94 &  H$_{6}$NaB$_{6}$ $\cdot$ 2H$_{2}$O(\#39376)  & 7 & & \multicolumn{1}{>{\columncolor{magenta}}c}{199} & \blue{K$_{2}$Sn$_{2}$O$_{3}$(\#40463)} & 104\\
95 & H$_{4}$Ca$_{2}$AsF$_{13}$(\#415156) & 15 & &  \multicolumn{1}{>{\columncolor{magenta}}c}{208} & Zn$_{3}$As$_{2}$(\#24486) & 9\\
96 & \blue{m-Cu$_{2}$S(\#16550)} & 105 & &   \multicolumn{1}{>{\columncolor{magenta}}c}{210} & H$_{6}$TeO$_{6}$(\#16435) & 7\\
98 & \blue{CdAs$_{2}$(\#16037)}  & 25  & & \multicolumn{1}{>{\columncolor{magenta}}c}{212} & \blue{Li$_{2}$Pd$_{3}$B(\#84931)} & 152\\
144 & \blue{LaBSiO$_{5}$(\#39756)}  & 89  & & \multicolumn{1}{>{\columncolor{magenta}}c}{213}  & \blue{Mg$_{3}$Ru$_{2}$(\#260022)} & 221 \\
145 & BiB$_{2}$O$_{4}$F(\#172481)   & 32 & & \multicolumn{1}{>{\columncolor{magenta}}c}{214} & \blue{Ag$_{3}$Se$_{2}$Au(\#171959)} & 33\\
151 & DyAl$_{3}$Cl$_{12}$(\#65975)& 48 & & & & \\
\hline
\end{tabular}
\end{table}
\end{center}

\begin{table}
\caption{\label{nonsymChiral} \textbf{Representative topological chiral crystals with Kramers-Weyl fermions in nonsymmorphic chiral space groups.} We list all 41 nonsymmorphic chiral space groups and, where possible, material candidates from the inorganic crystal structure database (ICSD)~\cite{ICSD}.  All the compounds listed in the table host Kramers-Weyl fermions. Among them, the compounds labeled in blue are (semi-)metals or small-gap insulators with clean Fermi surfaces and Kramers-Weyl fermions near the Fermi level; their electronic structures are shown in Supplemental Information \textbf{J}. The numbers in brackets are the ICSD collection codes. The space groups labeled in magenta may also host additional unconventional four-fold-degenerate chiral fermions at time-reversal-invariant momenta (TRIMs) with little groups isomorphic to chiral point groups.  Space groups 198 and 199 have TRIM-point little groups isomorphic to chiral point group $23$ ($T$), and may thus host the unconventional chiral fermion identified in Refs.~\cite{RhSi,CoSi}.  Space groups 208, 210, 212, 213, and 214 have TRIM-point little groups isomorphic to chiral point group $432$ ($O$), and may thus host the unconventional spin-3/2 chiral fermion identified in Ref.~\cite{NewF}.}
\end{table}

\clearpage
\begin{figure}[t]
\includegraphics[width=13cm]{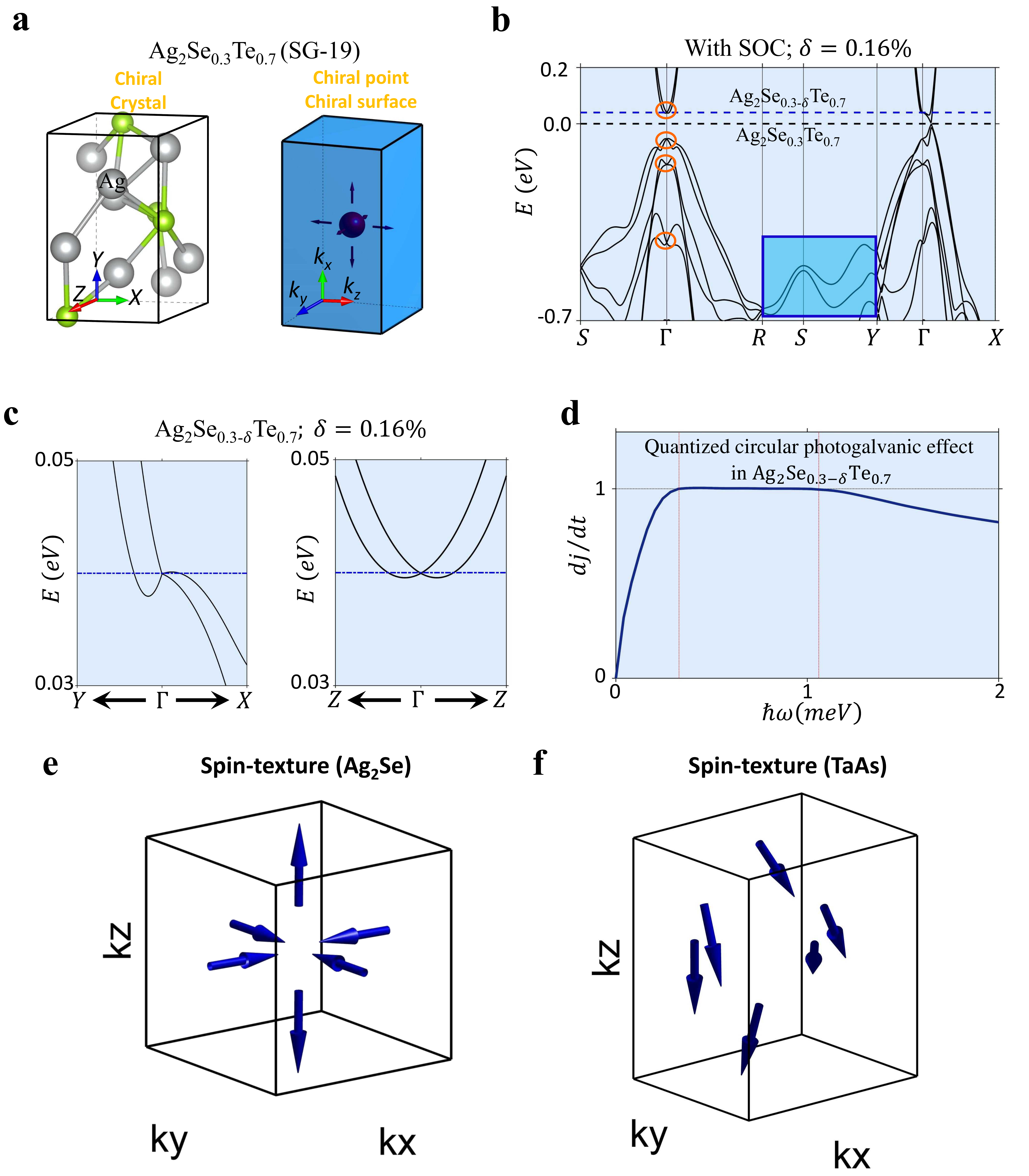}
\caption{{\bf Quantized circular photogalvanic current and real spin-momentum locking.}
({\bf a}) The left panel shows the chiral crystal of Ag$_2$Se$_{0.3}$Te$_{0.7}$. The right panel illustrates the chiral fermions in Ag$_2$Se$_{0.3}$Te$_{0.7}$, where the $\Gamma$ point allows an isolated Kramers-Weyl node and the BZ boundaries feature a nodal surface with nonzero chiral charge (SM \textbf{E,F}). ({\bf b}) The band structure of Ag$_2$Se$_{0.3}$Te$_{0.7}$. The Kramers-Weyl fermions at $\Gamma$ are highlighted by the orange circles. The chiral nodal surfaces are indicated by the blue boxes. The Fermi levels of undoped (Ag$_2$Se$_{0.3}$Te$_{0.7}$) and slightly electron-doped (Ag$_2$Se$_{0.3-\delta}$Te$_{0.7}$ ($\delta$=0.16\%)) samples are indicated by the black and blue dashed lines. ({\bf c}) Bands in the vicinity of the $\Gamma$ point.
({\bf d}) Quantized circular photogalvanic current induced by the Kramers-Weyl fermions in Ag$_2$Se$_{0.3-\delta}$Te$_{0.7}$ ($\delta=0.16\%$). The current rate saturates at the expected quantized value for a $|C|=1$ Weyl fermion for photon energies $0.3$ meV $\lesssim\hbar\omega\lesssim1.1$ meV. ({\bf e,f}) Real spin along the three principle axes for the Kramers-Weyl node near the Fermi level in Ag$_2$Se and a conventional Weyl node in TaAs. }
\label{Fig3}
\end{figure}

\newpage

\clearpage
\textbf{
\begin{center}
{\large \underline{Supplementary Materials:} \\Universal Topological Electronic Properties of Nonmagnetic Chiral
Crystals}
\end{center}
}

\vspace{0.2cm}

\begin{center}

\end{center}

\vspace{0.25cm}

\textbf{
\begin{center}
{\large This file includes:\\}
\end{center}
}
\vspace{0.45cm}
\textbf{
\begin{tabular}{l l}
\underline{SI A.} & Computational Methods\\
\underline{SI B.} & Group Theory for Kramers-Weyl Fermions\\
\underline{SI C.} & Allowed Degeneracies in the Presence of $\mathcal{C}_n$ Rotation Symmetries \\
\underline{SI D.} & Kramers-Weyl Nodes have an Odd Chern Number \\
\underline{SI E.} & Nodal Degeneracies in Nonsymmorphic Chiral Space Groups \\
\underline{SI F.} & Topology of Nodal Surfaces in Nonsymmorphic Chiral Space Groups \\
\underline{SI G.} & Kramers-Weyl Nodes in Achiral Space Groups \\
\underline{SI H.} & Fermi Surface Analysis of Kramers-Weyl Materials \\
\underline{SI I.} & Fermi Arcs in Kramers-Weyl Metals\\
\underline{SI J.} & Additional Kramers-Weyl Materials \\
\underline{SI K.} & Weyl Nodes Distribution in AgBi(Cr$_2$O$_7$)$_2$ \\
\underline{SI L.} & Theory for Novel Spin Texture of Kramers-Weyl Fermions \\
\underline{SI M.} & Calculated Spin Texture near Kramers-Weyl Points \\
\underline{SI N.} & Nonzero Chern Numbers of all Point Degeneracies in Chiral Crystals \\
\underline{SI O.} &  Comparison of Kramers-Weyl Fermions and Other Chiral Fermions\\
\end{tabular}
}

\newpage
\clearpage
\section*{SM A. Computational Methods}
We performed first-principles calculations within the density functional theory (DFT) framework using the projector augmented wave method \cite{dft2} as implemented in the VASP~\cite{dft3} package and the full-potential augmented plane-wave method as implemented in the package Wien2k \cite{dft4}. The generalized gradient approximation (GGA) was used \cite{dft5}. The lattice constants for the materials examined in this paper were obtained from the ICSD~\cite{ICSD}. To calculate the surface states of AgBi(Cr$_2$O$_7$)$_2$, Wannier functions were generated using the $d$ orbitals of Ag, the $p$ and $d$ orbitals of Cr, and the $p$ orbitals of O. The surface states were calculated for a semi-infinite slab by the iterative Green's function method.

In order to calculate the circular photogalvanic current in Ag$_2$Se$_{0.3}$Te$_{0.7}$ (SG~19), we generated the Wannier functions of Ag$_2$Se (SG~19) and Ag$_2$Te (SG~19) using the $s$ and $d$ orbitals of Ag and the $p$ orbitals of Se (Te). The electronic structure of Ag$_2$Se$_{0.3}$Te$_{0.7}$ (SG~19) was calculated by a linear interpolation between tight-binding model matrix elements of Ag$_2$Se (SG 19) and Ag$_2$Te (SG~19). We calculated the multiband gyrotropic tensor $\beta_{ij}(\omega)$ as described in Ref.~\cite{Q_photo_Cur} to obtain the circular photogalvanic effect photocurrent rate:
\begin{equation}
\beta_{ij}(\omega)=\frac{i\pi e^{3}}{\hbar}\epsilon_{jkl}\sum_{\bold{k},n,m}f^{\bold{k}}_{nm}\Delta^{i}_{\bold{k},nm}r^{\bold{k}}_{k,nm} r^{l}_{\bold{k},mn}\delta(\hbar\omega - E_{\bold{k},mn}).
\end{equation}

\section*{SI B. Group Theory for Kramers-Weyl Fermions}

Here, we provide a more formal definition of a Kramers-Weyl node in the language of group theory and irreducible representations.  Kramers-Weyl fermions are defined as two-dimensional, double-valued irreducible corepresentations of little groups that are isomorphic to the time-reversal-symmetric chiral point groups.  The chiral, or enantiomorphic, point groups are defined as those lacking rotoinversions~\cite{Q_photo_Cur,ChiralPoint}, and are therefore given by a short list: $1,2,3,4,6,222,422,622,32,23,$ and $432$.

It has been shown in previous works~\cite{WiederLayers,QuantumChemistry} that bands in strong-spin-orbit systems must be singly-degenerate when dispersing from $\Gamma$ in crystals with 3D symmetries, and therefore the two-dimensional irreducible corepresentations presented here will always have well-defined Chern numbers.  In the cases of nonsymmorphic groups, the little groups at the zone-edge TRIMs can contain projective representations modified by fractional lattice translations, and will therefore no longer be isomorphic to those at $\Gamma$. They thus will no longer be generically isomorphic to point groups.  However, we can state that in symmorphic chiral crystals, the little groups at all of the TRIMs will be isomorphic to time-reversal-symmetric chiral point groups, and thus will be able to host Kramers-Weyls fermions (though we note that in body- and face-centered crystals, some of the TRIMs may be isomorphic to different chiral point groups.

We can use Bradley and Cracknell~\cite{BigBook} to list the abstract groups and irreducible corepresentations that describe Kramers-Weyl fermions.  They are listed in Table~\ref{tab:corep}, obtained by considering the allowed irreducible corepresentations at $\Gamma$ for crystals in space groups which modulo translations are isomorphic to the chiral point groups.  We list a point group under two different naming conventions, give the simplest space group that generates crystals in this point group, and list the abstract groups and irreducible corepresentations of the Kramers-Weyl node for each of the chiral point groups.

We can also use the results of SI \textbf{C} and Ref.~\cite{Tsirkin:2017lfh} to determine the magnitudes of the Chern numbers $|C|$ of the Kramers-Weyl points described by the corepresentations listed in Table~\ref{tab:corep}.  Kramers-Weyl fermions with Chern numbers $|C|=3$ must have $C_{3}$ eigenvalues of $-1$, or $C_{6}$ eigenvalues of $\pm i$; all other chiral two-fold degeneracies with well-defined rotation eigenvalues must have Chern numbers $|C|=1$.  Therefore, considering the action of time-reversal symmetry on the eigenvalues, we search for corepresentations with either:
\begin{equation}
\chi_{\tilde{\rho}}\left(C_{3}\right)=-2\text{ or }\chi_{\tilde{\rho}}\left(C_{6}\right)=0,
\end{equation}
where $\chi_{\tilde{\rho}}(g)$ is the character of the point group element $g$ for the corepresentation $\tilde{\rho}$.

\begin{table}
\begin{centering}
\begin{tabular}{ |c|c|c|c|c| }
\hline
\multicolumn{5}{|c|}{Irreducible Corepresentations of Kramers-Weyl Fermions} \\
\hline
Point Group & Point Group & Simplest SG & Abstract Group and   & Magnitude of \\
  Name 1 & Name 2 & & Corepresentations under $\mathcal{T}$ & Chern Number ($|C|$) \\
 \hline
$1$ & $C_{1}$ & 1 & $G^{1}_{2}:\ R_{2}R_{2}$ & 1 \\
\hline
$2$ & $C_{2}$ & 3 & $G^{1}_{4}:\ R_{2}R_{4}$ & 1 \\
\hline
$222$ & $D_{2}$ & 16 & $G^{5}_{8}:\ R_{5}$ & 1 \\
\hline
$4$ & $C_{4}$ & 75 & $G^{1}_{8}:\ R_{2}R_{8},\ R_{4}R_{6}$ & 1\\
\hline
$422$ & $D_{4}$ & 89 & $G^{14}_{16}:\ R_{6},\ R_{7}$ & 1 \\
\hline
$3$ & $C_{3}$ &  143 & $G_{6}^{1}:\ R_{2}R_{6}$ & 1 \\
 & & & $G_{6}^{1}:\ R_{4}R_{4}$ & 3 \\
\hline
$32$ & $D_{3}$ & 149 & $G^{4}_{12}:\ R_{3}R_{4}$ & 3 \\
 & & & $G^{4}_{12}:\ R_{6}$ & 1 \\
\hline
$6$ & $C_{6}$ & 168 & $G_{12}^{1}:\ R_{2}R_{12},\ R_{6}R_{8}$ & 1  \\
 & & & $G_{12}^{1}:\ R_{4}R_{10}$ & 3 \\
\hline
$622$ & $D_{6}$ & 177 & $G_{24}^{11}:\ R_{7}$ & 3 \\
 & & & $G_{24}^{11}:\ R_{8},\ R_{9}$ & 1 \\
\hline
$23$ & $T$ & 195 & $G_{24}^{9}:\ R_{4}$ & 1 \\
\hline
$432$ & $O$ & 207 & $G_{48}^{10}:\ R_{4},\ R_{5}$ & 1 \\
\hline
\end{tabular}
\caption{The irreducible corepresentations of Kramers-Weyl fermions as labeled in Ref.~\cite{BigBook}.}
\label{tab:corep}
\end{centering}
\end{table}

\section*{SI C. Allowed Degeneracies in the Presence of $\mathcal{C}_n$ Rotation Symmetries}
\label{sec:evals}

In this section we study whether rotational symmetry $\mathcal{C}_n$, $n=2,3,4,6$ can enforce Kramers-Weyl nodes with chiral charge $|C|>1$. In the absence of time-reversal symmetry, generic Weyl nodes located on a $\mathcal{C}_n$ rotation axis were studied in Ref.~\cite{Multi.Weyl}. It was found that $\mathcal{C}_n$ with $n=2,3$ can only protect Weyl nodes with $C=\pm1$, $\mathcal{C}_4$ symmetry can protect Weyl nodes with both $C=\pm1$ and $C=\pm2$, and $\mathcal{C}_6$ symmetry can protect Weyl nodes with $C=\pm1$, $C=\pm2$, and $C=\pm3$. According to these findings, we in this section analyze cases with all values of $n$ to determine which are compatible with time-reversal symmetry and, therefore, are allowed at a Kramers-Weyl node.

To determine the possible kinds of degeneracies, we make use of the general classification scheme developed in Ref.~\cite{Multi.Weyl}: we first note that for a spinless case, the eigenvalues of $\mathcal{C}_n$ in the filled subspace are taken out of the set of the $n$-th roots of $+1$. In addition, time-reversal symmetry forces them to either be real or to come in complex-conjugated pairs. The same holds for the case of spinful bands, where the eigenvalues are further split and taken out of the set of the $n$-th roots of $-1$. Furthermore, taking every spinless eigenvalue $\chi$ of $\mathcal{C}_n$ belonging to a certain band that splits when introducing spin, we can uniquely assign two spinful eigenvalues for the split bands, which are determined by $u=\chi e^{\pm i\pi/n}$. The nature of the crossing between spinful bands is now determined by the fraction $u_{\mathrm{c}} /u_{\mathrm{v}}$, where $u_{\mathrm{c}}$ and $u_{\mathrm{v}}$ are the time-reversal paired eigenvalues. From this fraction, the absolute value of the chiral charge $|C|$ can be read off from Tab.~I in Ref.~\cite{Multi.Weyl}, see Tab.~\ref{tab:c4} and~\ref{tab:c6}.

From the content of this table, we can conclude that time-reversal symmetry in combination with $\mathcal{C}_4$ symmetry allows only for $C=\pm{1}$ Weyl nodes, while its combination with $\mathcal{C}_6$ symmetry is additionally compatible with $C=\pm{3}$ Weyl nodes. In the latter case, a spinless Kramers-Weyl node $C=\pm1$ at the Kramers point is split by spin-orbit coupling into two Weyl nodes, one with $C=\pm1$, and another with $C=\pm3$. An analysis of the irreducible representations in chiral space groups with $\mathcal{C}_{6}$ symmetry, such as in the simple example of SG 168 $P6$, confirms the allowed presence of both $C=|1|$ and $C=|3|$ Kramers-Weyl nodes~\cite{BigBook}. Note that as pointed out by Tsirkin \emph{et al.} in Ref.~\cite{Tsirkin:2017lfh}, in the case of $\mathcal{C}_{3}$ symmetry there is one additional possibility, which is not captured by restricting the classification of Ref.~\cite{Multi.Weyl} to time-reversal-symmetric systems: when a Kramers pair of bands has a real spinful $\mathcal{C}_{3}$ eigenvalue, it may only cross in a Weyl node with chiral charge $C = \pm 3$.

\begin{table}[h]
  \begin{tabular}{ | l | c | c  | c | c |}
    \hline
    $\mathcal{C}_4$ 	& spinless	 eigenvalues	 	& spinful eigenvalues			  						& $u_{\mathrm{c}} /u_{\mathrm{v}}$ 	&$|C|$	\\ \hline
     		& +1 			& $\{e^{i \frac{\pi}{4}}$,  $e^{-i \frac{\pi}{4}}$\}		& $\pm i$			&$1$		\\ \hline
     		& -1 				& $\{-e^{i \frac{\pi}{4}}$,  $-e^{-i \frac{\pi}{4}}$\}		& $\pm i$			&$1$		\\ \hline
     		& $\{+i, -i\}	$		& $(\{e^{i \frac{3\pi}{4}},e^{-i \frac{3\pi}{4}}\}$, $\{e^{i \frac{\pi}{4}},e^{-i \frac{\pi}{4}}\})$		& $(\pm i,\pm i)$		& $(1,1)$	\\
    \hline
  \end{tabular}
\caption{Properties of allowed $\mathcal{C}_4$ band crossings at Kramers points. The first two lines each correspond to a single, spin-degenerate band, which splits under spin-orbit coupling into two bands that cross at a Kramers-Weyl node with unit chiral charge. The last line describes the splitting of two spin-degenerate bands, forming a spinless Weyl cone at a TRIM into four bands, which exhibit two Kramers crossings of unit chiral charge, respectively.}
\label{tab:c4}
\end{table}

\begin{table}[h]
  \begin{tabular}{ | l | c | c  | c | c |}
    \hline
    $\mathcal{C}_6$ 	& spinless eigenvalues		& spinful eigenvalues	 							& $u_{\mathrm{c}} /u_{\mathrm{v}}$ 			&$|C|$	\\ \hline
     		& +1 				& \{$e^{i \frac{\pi}{6}}$, $e^{-i \frac{\pi}{6}}$\}			& $e^{\pm i \frac{\pi}{3}}$		&$1$		\\ \hline
     		& -1 					& \{$-e^{i \frac{\pi}{6}}$, $-e^{-i \frac{\pi}{6}}$\}	& $e^{\pm i \frac{\pi}{3}}$		&$1$		\\ \hline
     		& \{$e^{i \frac{\pi}{3}},e^{-i \frac{\pi}{3}}\}$		 &(\{$e^{i \frac{3\pi}{6}},e^{-i \frac{3\pi}{6}}\}$,$\{e^{i \frac{\pi}{6}},e^{-i \frac{\pi}{6}}\})$	& $(-1,e^{\pm i \frac{\pi}{3}})$					&$(3,1)$		\\ \hline
     		& \{$e^{i \frac{2\pi}{3}},e^{-i \frac{2\pi}{3}}\}$	 &(\{$e^{i \frac{5\pi}{6}},e^{-i \frac{5\pi}{6}}\}$,$\{e^{i \frac{3\pi}{6}},e^{-i \frac{3\pi}{6}}\})$	& $(e^{\pm i \frac{\pi}{3}},-1)$					&$(1,3)$		\\
    \hline
  \end{tabular}
\caption{Properties of allowed $\mathcal{C}_6$ band crossings at Kramers points. The first two lines each correspond to a single, spin-degenerate band, which splits under spin-orbit coupling into two bands that cross at a Kramers-Weyl nodes with unit chiral charge. The third and fourth lines both describe the splitting of two spin-degenerate bands crossing at a TRIM into four bands, which exhibit two different Kramers crossings: one of unit chiral charge and another of chiral charge $3$.}
\label{tab:c6}
\end{table}

\section*{SI D. Kramers-Weyl Nodes have an Odd Chern Number}

In this section we show that for a time-reversal-invariant two-band Hamiltonian $H(\boldsymbol{k}) = \boldsymbol{d}(\boldsymbol{k}) \cdot \boldsymbol{\hat{\sigma}}$ with $\mathcal{T}^2=-1$, the Chern number of a (Fermi) surface enclosing a TRIM is always odd. Here, the elements of vector $\boldsymbol{\hat{\sigma}}$ are the three Pauli matrices and $\boldsymbol{k}$ lives in three-dimensional momentum space.
Time-reversal symmetry can be represented by $\mathcal{T} = \mathit{K} i \sigma^y$, which flips the spin $\mathcal{T}\boldsymbol{\hat{\sigma}}\mathcal{T}^{-1}=-\boldsymbol{\hat{\sigma}}$. Therefore, time-reversal symmetry of the Hamiltonian $\mathcal{T} H(\boldsymbol{k}) \mathcal{T}^{-1} = H(-\boldsymbol{k})$ implies $\boldsymbol{d}(\boldsymbol{k}) = - \boldsymbol{d}(-\boldsymbol{k})$.

Since the Chern number of a surface in momentum space is a topological quantity, we may choose our manifold of integration to be the unit sphere $S^2_{\boldsymbol{0}}$ centered at the origin (the TRIM) without loss of generality. The Chern number is then given by
\begin{equation}
C = \frac{1}{2\pi} \int_{S^2_{\boldsymbol{0}}} \mathrm{d}A,\qquad A = i\,\bra{u(\boldsymbol{k})} \mathrm{d} \ket{u(\boldsymbol{k})},
\end{equation}
where $\mathrm{d}$ is the exterior derivative and $\ket{u(\boldsymbol{k})}$ denotes the eigenvector of $H(\boldsymbol{k})$ belonging to its lower eigenvalue. Note that since the Hamiltonian is gapped on $S^2_{\boldsymbol{0}}$, we can take $|{\boldsymbol{d}(\boldsymbol{k} \in S^2_{\boldsymbol{0}})}| = 1$ without loss of generality.

Due to the antisymmetry of $\boldsymbol{d}(\boldsymbol{k})$ on this sphere, there will always be a singly-connected time-reversal-invariant closed loop $\mathcal{C}_{d^z=0} \subset S^2_{\boldsymbol{0}}$ of $\boldsymbol{k}$-points  on which $d^z(\boldsymbol{k})=0$. On the loop $\mathcal{C}_{d^z=0}$, we may then re-express the Hamiltonian as
\begin{equation}
H(\boldsymbol{k}) =  \left(\begin{matrix}0&e^{-i \Phi(\boldsymbol{k})} \\ e^{i \Phi(\boldsymbol{k})}&0\end{matrix} \right), \quad \Phi(\boldsymbol{k}) = \Phi(-\boldsymbol{k}) + (2 k + 1) \, \pi \text{, where } k \in \mathbb{Z},
\end{equation}
where the second equality follows from time-reversal symmetry. It is now straightforward to solve for $A = \frac{1}{2} \partial_\phi \Phi(\phi)$, where $\phi\in[0,2\pi]$ denotes an angle parametrizing $\mathcal{C}_{d^z=0}$.
For parallel transport around any closed loop on the Bloch sphere, we have the relation
\begin{equation}
\mathrm{exp}\left(
i \oint_{\mathcal{C}} A
\right)
=
\mathrm{exp}\left(
i \int_{\Lambda} \mathrm{d}A
\right),
\label{eq: Berry equivalence}
\end{equation}
where $\Lambda$ is a two-dimensional submanifold of the sphere and $\mathcal{C}$ is its boundary.

Since by time-reversal symmetry, the segments of the sphere separated by $\mathcal{C}_{d^z=0}$ yield the same contribution to the Chern number, we can use Eq.~\eqref{eq: Berry equivalence} to arrive at
\begin{equation}
C = 2 \left(\frac{1}{2\pi} \oint_{\mathcal{C}_{d^z=0}} A +m\right)= \frac{1}{2\pi} \int_0^{2 \pi} \mathrm{d}\phi \, \partial_\phi \Phi(\phi)+2m,\qquad m\in \mathbb{Z}.
\end{equation}

Note that the periodicity of $\Phi(\phi)$ implies $\Phi(2\pi) = \Phi(0) + \, 2\pi n$, where $n \in \mathbb{Z}$. Together with the constraint imposed by time-reversal symmetry above, we actually even gain the stronger constraint $\Phi(2\pi) = \Phi(0) + 2\pi (2 k + 1) $, where $k \in \mathbb{Z}$, i.e., $n$ is an odd integer. This implies that $C = 2(k+m) +1$ is odd and, therefore, that Kramers-Weyl nodes have an odd Chern number.

\section*{SI E.  Nodal Degeneracies in Nonsymmorphic Chiral Space Groups}

In this section we discuss the effects of nonsymmorphic symmetries in chiral space groups on the band structures near TRIM points at the BZ boundary. For a subset of nonsymmorphic chiral space groups, all Kramers degeneracies at the TRIMs will have $\mathcal{T}$-pinned Kramers-Weyl nodes. However, another subset will have Kramers degeneracies that form nodal planes, instead of isolated Weyl nodes due to the relationship between their screw rotation axes and time-reversal-symmetry.

\begin{figure*}[h]
\centering
\begin{center}
\includegraphics[width=16cm]{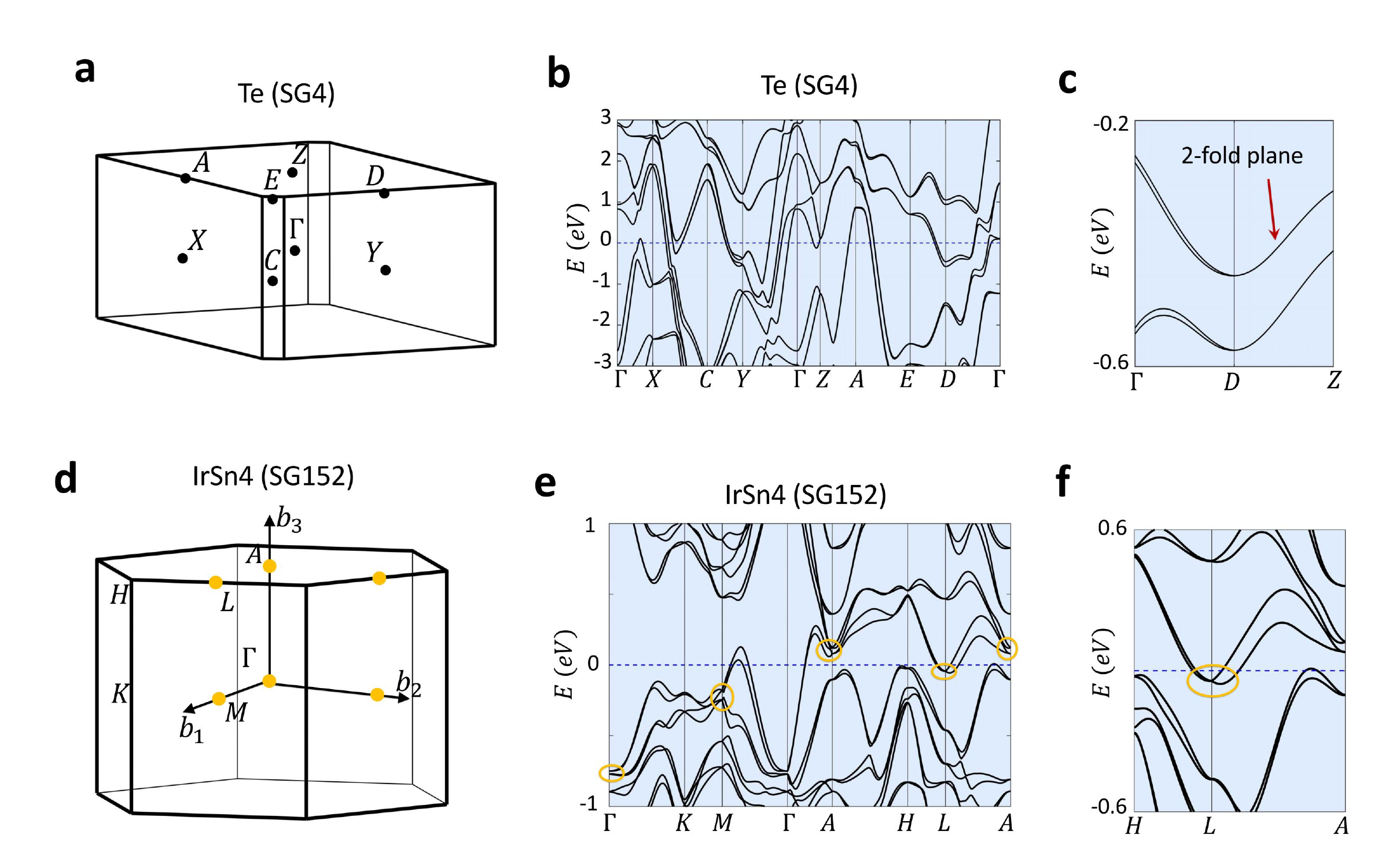}
\end{center}
\caption{\label{Fig1S} \textbf{Kramers-Weyl nodes in nonsymmorphic chiral space groups.} (\textbf{a}) The bulk BZ and TRIM points of Te in  SG 4. (\textbf{b}) The electronic band structure of Te along high-symmetry lines.  (\textbf{c}) Bands in the vicinity of $D$. Along $D-Z$, which lies in the $k_{z}=\pi$ plane, bands are two-fold degenerate. (\textbf{d}) The BZ and TRIM points of SG 152. (\textbf{e}) The band structure of IrSn$_4$ in SG 152 along the high symmetry lines $H-L$ and $L-A$. (\textbf{f}) Bands passing through the $L$ point at the zone boundary confirm that this TRIM hosts Kramers-Weyl fermions.}
\end{figure*}

The zone-boundary degeneracy structure in nonsymmorphic chiral crystals is strongly affected by the presence of nodal planes, which can form when certain screw rotation symmetries combine with time-reversal symmetry.  For a crystal symmetry $\Pi$ to combine with time-reversal to protect a two-fold degeneracy away from a TRIM, $\mathcal{T}\times \Pi$ must return $\bs{k}$ to itself and square to $-1$ for that value of $\bs{k}$. Thus, in a $\mathcal{T}$-symmetric crystal with this $\Pi$ symmetry, bands will be at least two-fold degenerate at $\mathcal{T}\times \Pi$-invariant $\bs{k}$-points in the Brillouin zone (BZ). The modified algebra of the screw rotations at $k_{i}=\pi$ can also be understood by noting that the little groups of those TRIMs are no longer isomorphic to those at $\Gamma$, as they were in SG 16, due to the projective effects of the fractional lattice translations on the commutation relations of the symmetry representations~\cite{NS-3DDirac}. To illustrate this more clearly, we use Te and IrSn$_4$ as real crystal examples (Figs.~\ref{Fig1S}\textbf{a-c}). The crystal lattice of Te is represented by the nonsymmorphic chiral space group 4 (SG 4) and, as a result, has a $s_{z}=\{\mathcal{C}_{2z}|00\frac{1}{2}$\} screw-rotation symmetry. Therefore, for the case of Te, $\mathcal{C}_{2z}$ sends $(k_{x}, k_{y}, k_{z})\rightarrow(-k_{x}, -k_{y}, k_{z})$ such that $s_{z}\times \mathcal{T}$ is a valid symmetry operation that maps a $k$-point to itself on the $k_{z}=0, \pi$ planes.  When acting on a Bloch eigenstate of the Hamiltonian, $(s_{z}\mathcal{T})^{2}=e^{-i k_z}$, a two-fold degeneracy is enforced on the entire $k_{z}=\pi$ plane. Thus, bands on the $k_{z} = \pi$ plane are two-fold degenerate in Te (SG 4). Furthermore, because points $Z$, $A$, $D$, and $E$ lie on the nodal plane, only the isolated two-fold degeneracies at $\Gamma$, $X$, $Y$, and $C$ are Kramers-Weyl nodes.

Conversely, three-fold screw rotations, which can only contain $1/3$ or $2/3$ fractional translations, cannot satisfy this algebraic constraint, and thus cannot obscure Kramers-Weyl points with nodal planes.  In order to enforce $(s_{3z}\mathcal{T})^{2}$=$-1$, $s_{3z}$ must be capable of squaring to +1 (or should attain that value at least when being raised to an even power). However, as in these cases only $C_{3z}^{3}=e^{ik_{z}}$, there is no way for it to be combined with $\mathcal{T}$ to lead to a local Kramers theorem. Note that this stands in contrast to a $6_{3}$ screw rotation $s_{6z}$=\{$C_{3z}|00\frac{1}{2}$\}, which can satisfy the required property.  Consider IrSn$_4$ (SG 152) as an example (Figs.~\ref{Fig1S}\textbf{d-f}), which has a $3_{1}$ screw rotation $s_{3z}$=\{$C_{3z}|00\frac{1}{3}$\}, and nevertheless hosts Kramers-Weyl nodes at all of its TRIM points.

We can exhaustively deduce which of the possible screw axes allowed in 3D crystals support nodal planes when combined with time-reversal symmetry.  Since nodal planes are enforced by anti-unitary symmetries which map every point of the plane to itself and square to $-1$, for a screw rotation symmetry $s_{n,i}$, $i = x,y,z$, we require that taken to some power $p$, $s_{n,i}^p$ acts as a $\mathcal{C}_2$ rotation on the in-plane momenta. This ensures that the $k_i = \pi$ plane, which is left invariant by $\mathcal{T}$, can in principle be subject to a local constraint which derives from the nonsymmorphic symmetry, and limits the set of available screw rotations to those with $n=2,4,6$. A second criterion we have to impose is $(s_{n,i}^p\mathcal{T})^{2} = e^{\mathrm{i} k_i q}$, where $q$ is an odd integer, so that $(s_{i,n}^p\mathcal{T})^{2} = -1$ on the $k_i = \pi$ plane enforces a local Kramers degeneracy. We therefore conclude that nodal planes can only be enforced by under the following screw symmetries: a two-fold rotation together with a $1/2$ lattice translation ($2_{1}$), a four-fold rotation with a $1/4$ translation ($4_{1}$), a sixfold rotation with a $1/6$ translation ($6_{1}$), as well as a sixfold rotation with a $1/2$ translation ($6_{3}$).  We also observe that for all of these screw space group elements $g$, $g^{p}\equiv 2_{1}$, such that we arrive at a broad condition that in time-reversal-symmetric, double-valued (strong-SOC) space groups $G$, nodal planes are supported if $2_{1}\in G$.

Consider as a final example the four-fold screw $s_{4z}=\{\mathcal{C}_{4z}|\frac{1}{4}\frac{3}{4}\frac{1}{4}\}$ in SG 214, which involves a quarter translation along the $z$ direction, along with fractional transversal translations. We can form the operator
\begin{equation}
\begin{aligned}
s_{4z}^2 &= e^{-\mathrm{i} \left( \frac{\hat{k}_x}{4} +  \frac{3 \hat{k}_y}{4} +  \frac{\hat{k}_z}{4} \right)} \mathcal{C}_{4z} \,
e^{-\mathrm{i} \left( \frac{\hat{k}_x}{4} +  \frac{3 \hat{k}_y}{4} +  \frac{\hat{k}_z}{4} \right)} \mathcal{C}_{4z} \\
&= e^{-\mathrm{i} \left( \hat{k}_x +  \frac{\hat{k}_y}{2} +  \frac{\hat{k}_z}{2} \right)} \mathcal{C}_{2z},
\end{aligned}
\end{equation}
where at this level $\hat{k}_i$ are still operators which are affected by crystal symmetry operations, such as $\mathcal{C}_{n}$ rotations, as well as by $\mathcal{T}$, where we note in particular that $\mathcal{T} \hat{k}_i \mathcal{T}^{-1} = - \hat{k}_i$.  This implies
\begin{equation}
\begin{aligned}
(s_{4z}^2 \mathcal{T})^2 &= e^{-\mathrm{i} \left( \hat{k}_x +  \frac{\hat{k}_y}{2} +  \frac{\hat{k}_z}{2} \right)} \mathcal{C}_{2z} \mathcal{T} \,
e^{-\mathrm{i} \left( \hat{k}_x +  \frac{\hat{k}_y}{2} +  \frac{\hat{k}_z}{2} \right)} \mathcal{C}_{2z} \mathcal{T} \\
&= e^{-\mathrm{i} k_z} \mathcal{C}_{2z}^2 \mathcal{T}^2 = e^{-\mathrm{i} k_z},
\end{aligned}
\end{equation}
enforcing a Kramers degeneracy everywhere on the $k_z = \pi$ plane. It becomes clear that transversal lattice translations always cancel out. Note also that this calculation is equally valid in the limits of weak and strong SOC.  We note that the focus of this section is purely on degeneracy, and not topology.  We find in SI \textbf{F} that, in fact, these nodal planes may still be topologically nontrivial.

\section*{SI F. Topology of Nodal Surfaces in Nonsymmorphic Chiral Space Groups}

\begin{figure*}[t]
\begin{center}
\includegraphics[width= \textwidth,page=1]{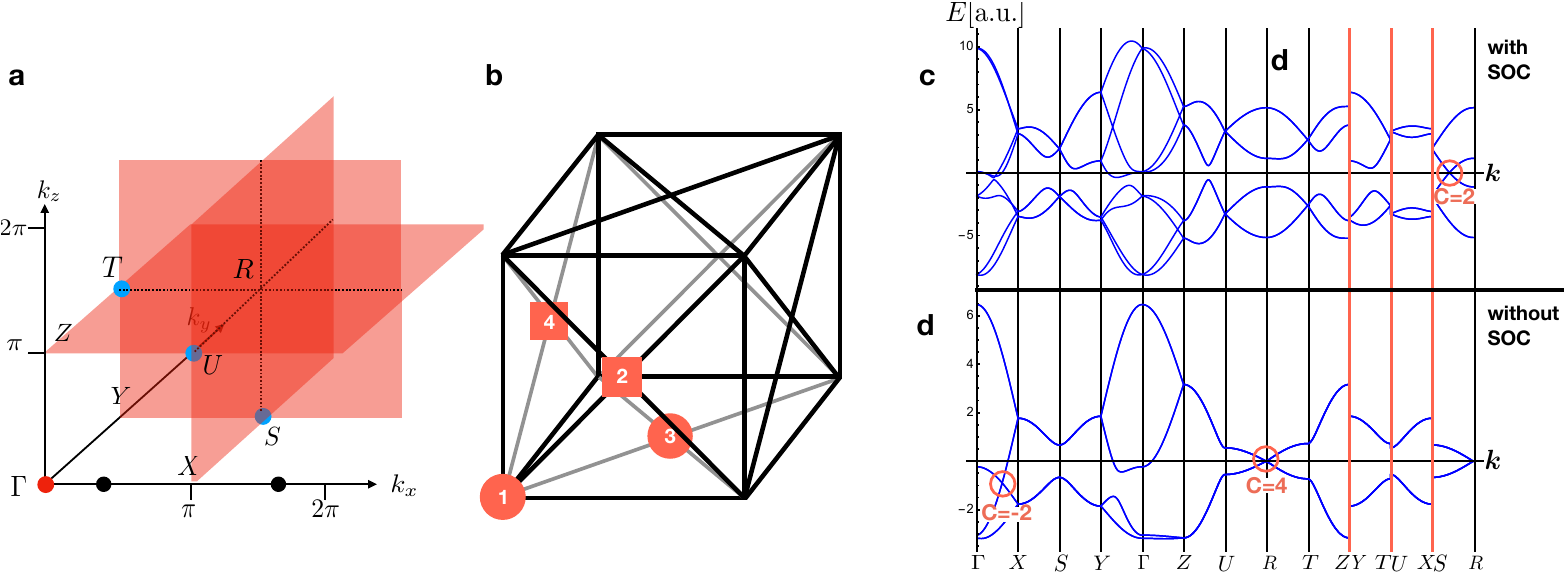}
\caption{
\textbf{Nodal surfaces in SG 19.}
(\textbf{a}) BZ for SG 19, with a nodal surface (red) at fillings $\nu\in 2+4\mathbb{Z}$, on which at the zone edges all bands are two-fold degenerate.  At the same fillings, the minimal connectivity of SG 19 also displays a two-fold-degenerate Kramers-Weyl fermion (red dot) at $\Gamma$.  Blue spheres indicate the four-fold-degenerate points that occur at the intersections of two nodal planes.  (\textbf{b}) Unit cell of the lattice on which the tight-binding Hamiltonian~\eqref{eq: SMsg19nodalsurfaceH} is defined. (\textbf{c})  Band structure of the tight-binding Hamiltonian~\eqref{eq: SMsg19nodalsurfaceH} incorporating the effects of SOC, plotted with parameter values $a_1 \cdots a_{22} =$ (0.593237, 0.258583, 0.547374, 0.27201, 0.596372, 0.959641, 0.945241, 0.191225, 0.557652, 0.583338, 0.461802, 0.00999746, 0.862848, 0.592332, 0.781321, 0.791785, 0.760452, 0.394747, 0.860525, 0.145271, 0.25431, 0.308917). The fourfold degeneracies along the SR carry Chern number $C=2$; there are four other Weyl points with $C=-1$ in the $k_{y}=0$ plane that are not visible in this   plot.  (\textbf{d})  Band structure of Eq.~\eqref{eq: SMsg19nodalsurfaceH} with all of the SOC terms set to zero.  There is a spin-degenerate eightfold chiral fermion at $R$ with charge $C=-4$ and a pair of filling-enforced spinless double-Weyl points along $\Gamma X$, each with charge $C=2$, in agreement with the results of Refs.~\cite{manes,Adrian19,Balatsky19}.  Upon the reintroduction of SOC, all of these points split, but the system remains gapless at $\nu=4$.  This indicates that the four conventional spinful Weyl points at $k_{y}=0$ are in fact \emph{filling-enforced}~\cite{WPVZ,WiederLayers}; despite not being characterized by symmetry eigenvalues, they enforce the minimal band connectivity~\cite{WiederLayers,QuantumChemistry} of SG 19 \emph{through a BZ plane}.
}
\label{fig: SG19SM}
\end{center}
\end{figure*}

Here we discuss the topological properties of nodal surfaces in the BZs of materials with nonsymmorphic chiral space groups. We here focus on the case of SG 19, the space group of the Kramers-Weyl compound Ag$_2$Se presented in the main text. As stated at the end of Sec.~I, this space group has three orthogonal two-fold screw rotation symmetries
\begin{equation}
\label{eq: screwsSG19}
s_{x}=\{\mathcal{C}_{2x}|\frac{1}{2}\frac{1}{2}0\},\quad s_{y}=\{\mathcal{C}_{2y}|0\frac{1}{2}\frac{1}{2}\},\quad s_{z}=\{\mathcal{C}_{2z}|\frac{1}{2}0\frac{1}{2}\}.
\end{equation}
Acting on a Bloch state with momentum $\bs{k}$, these symmetries obey $(s_{i}\mathcal{T})^{2}=e^{-i k_i}$ and for each $i = x,y,z$, the anti-unitary symmetry $s_{i}\mathcal{T}$ leaves momenta in the $k_i = 0,\pi$ plane invariant. In particular, on the $k_i = \pi$ planes they square to $-1$ and by Kramers pairing therefore imply a two-fold degeneracy on all surfaces forming the BZ boundary (instead of just the time-reversal invariant momenta), which may thus be viewed as a symmetry enforced nodal surface that may carry a topological charge.  Figure~\ref{fig: SG19SM}\textbf{a} shows the location of the nodal surace in the BZ.

We note that at time-reversal symmetric momenta at which $k_i = \pi$ holds for exactly two $k_i$, the nonsymmorphic symmetries enforce a four-fold degeneracy. We can show this for, e.g., the point $\bs{k} = (\pi, \pi, 0)$ by studying the algebra
\begin{equation}
s_x s_y = - e^{i (k_y + k_z)} s_y s_x e^{i k_x} = - s_y s_x
\end{equation}
that holds when acting on a Bloch state at $\bs{k} = (\pi, \pi, 0)$.  At this point, time-reversal symmetry $\mathcal{T}$ commutes with both $s_y$ and $s_x$, which both square to $+1$ and mutually anticommute.  This algebra enforces a four-fold degeneracy of every band at $\bs{k} = (\pi, \pi, 0)$, as highlighted in Refs.~\cite{WiederLayers,WallpaperFermions}.   These TRIM points are labeled with blue spheres in Fig~\ref{fig: SG19SM}\textbf{a}.  At TRIMs for which only one or none of the $k_{i}=\pi$, the corresponding algebra is not satisfied and does not enforce higher degeneracies.

The nodal surfaces on the zone boundary in fact can also have a topological charge~\cite{NodalSurfaces}, which  is given by the Chern number on an enclosing surface. The procedure for calculating this number is similar to enclosing a Weyl point with a sphere and calculating the Chern number of the eigenstates on the sphere.  For a nodal plane, an example of such an enclosing surface can be formed by taking all of the points whose shortest distance to any point on the nodal surface is given by some infinitesimal number $\delta$.  As long as the spectrum on the enclosing surface remains gapped under smooth deformations of the Hamiltonian, its Chern number is fixed, yielding an integer topological invariant for the nodal surface.

When time-reversal symmetry is imposed, the chiral charge of the nodal plane in SG 19 can only take odd-integer values. To see this, note that the surface enclosing the BZ boundary at fillings $\nu=2,6$ may also be viewed as a surface enclosing the rest of the BZ interior, including the $\Gamma$ point, at which one Kramers-Weyl node resides [red dot in Fig.~\ref{fig: SG19SM}\textbf{a}]. Since these viewpoints only differ by an orientation reversal of the surface, the nodal surface Chern number has to be the opposite of that of the Weyl node at $\Gamma$, which is given by $C = \pm 1$, plus that of any other Weyl nodes in the BZ bulk. By time-reversal symmetry, such Weyl nodes in the BZ bulk may only appear in pairs with the same Chern numbers. For example, the black dots in Fig.~\ref{fig: SG19SM}\textbf{a} represent a pair of same charge Weyl points that are compatible with all the symmetries of SG 19 and time-reversal. In summary, in SG 19, because the Kramers-Weyl node at $\Gamma$ has a chiral charge $\chi_\Gamma=\pm1$, the nodal plane at the BZ has an odd-integer chiral charge $\chi_{\mathrm{p}}$, where $\chi_\Gamma-\chi_{\mathrm{p}}\in 2\mathbb{Z}$ is the total chiral charge of other Weyl points in the bulk of the BZ at the same filling as the Kramers-Weyl fermion at $\Gamma$.

We now present an explicit tight-binding model for SG 19 displaying the aforementioned surface topology. A particularly natural realization of the symmetries given in Eq.~\eqref{eq: screwsSG19} is given by a simple fcc lattice with unit cell as shown in Fig.~\ref{fig: SG19SM}\textbf{b}, where the nearest-neighbor sites are related not by pure lattice translations, but instead by screw symmetries, such that the bravais lattice itself is still orthorhombic. To construct a tight-binding Hamiltonian, we begin with a model that includes all possible nearest-neighbor hoppings and SOC couplings, which has $6 \times 4 \times 4 = 96$ free parameters, as for each of the ${4 \choose 2} = 6$ possible bond pairings there are four directions along which an electron can hop with a spin rotation given by one of the three Pauli spin matrices $\sigma_i$, $i = x,y,z$ or the identity $\mathbb{I}$. Enforcing time-reversal and nonsymmorphic symmetries, we arrive at a model with $22$ free parameters $a_1 \cdots a_{22}$ and a Bloch Hamiltonian
\begin{equation}
\label{eq: SMsg19nodalsurfaceH}
\begin{aligned}
H(\bs{k}) &= \begin{pmatrix}
0 & h_{12}(\bs{k}) & h_{13}(\bs{k}) & h_{14}(\bs{k})�\\
h_{12}(\bs{k})^\dagger & 0 & h_{23}(\bs{k}) & h_{24}(\bs{k}) \\
h_{13}(\bs{k})^\dagger & h_{23}(\bs{k})^\dagger & 0 & h_{34}(\bs{k}) \\
h_{14}(\bs{k})^\dagger & h_{24}(\bs{k})^\dagger & h_{34}(\bs{k})^\dagger & 0
\end{pmatrix},\\
h_{ij} (\bs{k}) &= h_{ij}^\mathrm{hop} (\bs{k}) \sigma_0 + \bs{h}_{ij}^\mathrm{SOC} (\bs{k}) \cdot \bs{\sigma},
\end{aligned}
\end{equation}
where the hopping matrix elements are given by
\begin{equation}
\begin{aligned}
h_{12}^\mathrm{hop} (\bs{k}) &= a_1 + a_2 e^{- i  k_x} + a_1 e^{- i  k_z} + a_2 e^{- i  (k_x + k_z)},\\
h_{14}^\mathrm{hop} (\bs{k}) &= a_3 + a_3 e^{- i  k_y} + a_4 e^{- i  k_z} + a_4 e^{- i  (k_y + k_z)},\\
h_{23}^\mathrm{hop} (\bs{k}) &= a_4 + a_4 e^{- i  k_y} + a_3 e^{+ i  k_z} + a_3 e^{- i  (k_y - k_z)},\\
h_{34}^\mathrm{hop} (\bs{k}) &= a_2 + a_1 e^{+ i  k_x} + a_2 e^{- i  k_z} + a_1 e^{+ i  (k_x - k_z)},\\
h_{13}^\mathrm{hop} (\bs{k}) &= a_5 + a_5 e^{- i  k_x} + a_6 e^{- i  k_y} + a_6 e^{- i  (k_x + k_y)},\\
h_{24}^\mathrm{hop} (\bs{k}) &= a_6 + a_6 e^{+ i  k_x} + a_5 e^{- i  k_y} + a_5 e^{+ i  (k_x - k_y)},
\end{aligned}
\end{equation}
and the spin-orbit couplings by
\begin{equation*}
\begin{aligned}
\bs{h}_{12}^\mathrm{SOC} (\bs{k}) &=  i  \begin{pmatrix}
a_{11} + a_{12} e^{- i  k_x} + a_{12} e^{- i  k_z} + a_{12} e^{- i  (k_x + k_z)} \\
a_{13} + a_{14} e^{- i  k_x} + a_{13} e^{- i  k_z} + a_{14} e^{- i  (k_x + k_z)} \\
a_{15} - a_{16} e^{- i  k_x} - a_{15} e^{- i  k_z} + a_{16} e^{- i  (k_x + k_z)}
\end{pmatrix},
\end{aligned}
\end{equation*}
\begin{equation*}
\begin{aligned}
\bs{h}_{14}^\mathrm{SOC} (\bs{k}) &=  i  \begin{pmatrix}
a_{17} + a_{17} e^{- i  k_y} + a_{18} e^{- i  k_z} + a_{18} e^{- i  (k_y + k_z)} \\
a_{19} - a_{19} e^{- i  k_y} - a_{20} e^{- i  k_z} + a_{20} e^{- i  (k_y + k_z)} \\
a_{21} + a_{21} e^{- i  k_y} + a_{22} e^{- i  k_z} + a_{22} e^{- i  (k_y + k_z)}
\end{pmatrix},
\end{aligned}
\end{equation*}
\begin{equation}
\begin{aligned}
\bs{h}_{23}^\mathrm{SOC} (\bs{k}) &=  i  \begin{pmatrix}
-a_{18} - a_{18} e^{- i  k_y} - a_{17} e^{+ i  k_z} - a_{17} e^{- i  (k_y - k_z)} \\
-a_{20} + a_{20} e^{- i  k_y} + a_{19} e^{+ i  k_z} - a_{19} e^{- i  (k_y - k_z)} \\
a_{22} + a_{22} e^{- i  k_y} + a_{21} e^{+ i  k_z} + a_{21} e^{- i  (k_y - k_z)}
\end{pmatrix},
\end{aligned}
\end{equation}
\begin{equation*}
\begin{aligned}
\bs{h}_{34}^\mathrm{SOC} (\bs{k}) &=  i  \begin{pmatrix}
a_{12} + a_{11} e^{+ i  k_x} + a_{12} e^{- i  k_z} + a_{11} e^{+ i  (k_x - k_z)} \\
-a_{14} - a_{13} e^{+ i  k_x} - a_{14} e^{- i  k_z} - a_{13} e^{+ i  (k_x - k_z)} \\
-a_{16} + a_{15} e^{+ i  k_x} + a_{16} e^{- i  k_z} - a_{15} e^{+ i  (k_x - k_z)}
\end{pmatrix},
\end{aligned}
\end{equation*}
\begin{equation*}
\begin{aligned}
\bs{h}_{13}^\mathrm{SOC} (\bs{k}) &=  i  \begin{pmatrix}
0 \\
a_{7} + a_{7} e^{- i  k_x} + a_{8} e^{- i  k_y} + a_{8} e^{- i  (k_x + k_y)} \\
a_{9} + a_{9} e^{- i  k_x} + a_{10} e^{- i  k_y} + a_{10} e^{- i  (k_x + k_y)}
\end{pmatrix},
\end{aligned}
\end{equation*}
\begin{equation*}
\begin{aligned}
\bs{h}_{24}^\mathrm{SOC} (\bs{k}) &=  i  \begin{pmatrix}
0 \\
-a_{8} - a_{8} e^{ i  k_x} - a_{7} e^{- i  k_y} - a_{7} e^{+ i  (k_x - k_y)} \\
a_{10} + a_{10} e^{ i  k_x} + a_{9} e^{- i  k_y} + a_{9} e^{+ i  (k_x - k_y)}
\end{pmatrix}. \\
\end{aligned}
\end{equation*}

Calculating the Chern number of this Hamiltonian on a surface enclosing the BZ boundary formed by the planes $\{(k_x,k_y,\pi \pm \delta),(k_y,\pi \pm \delta,k_x),(\pi \pm \delta,k_x,k_y)\}$, $k_x, k_y \in (-\pi + \delta, \pi - \delta)$ with $\delta = 0.1$ for the generic model parameters quoted in the caption of Fig.~\ref{fig: SG19SM}, we confirm the above arguments and obtain $C = 1$.

In addition to Kramers-Weyl nodes at its TRIM points at fillings $\nu=2,6$, Eq.~\eqref{eq: SMsg19nodalsurfaceH} exhibits additional chiral fermions at half-filling, or $\nu=4$.  In Figs.~\ref{fig: SG19SM}~\textbf{c} and~\textbf{d} we plot the bands of this model with and without the SOC terms, respectively.  When SOC is negligible, this model exhibits a spin-degenerate eightfold chiral fermion at $R$ with charge $C=-4$ and a pair of filling-enforced spinless double-Weyl points along $\Gamma X$, each with charge $C=2$, in agreement with the results of Refs.~\cite{manes,Adrian19,Balatsky19}.  When SOC is reintroduced, these points split, but the filling-enforced gaplessness at $\nu=4$ is still preserved by the spinful symmetries of SG 19~\cite{WPVZ,WiederLayers}.  The chiral fermion at $R$ splits into two nonsymmorphic, zone-edge double-Weyl points~\cite{BarryPrivate,BarryPrep} (SI~\textbf{N}), each with charge $C=-2$.  More interestingly, the moveable-but-unremovable spinless double-Weyl points along $\Gamma X$ split into four conventional $C=1$ Weyl points in the $k_{y}=0$ plane.  To understand this splitting, we observe that the conservation of chiral charge requires that the sum of the Chern numbers of the nodal degeneracies in the vicinity of previous spinless double-Weyl points is $+2$, and note that the spinful screw symmetries cannot preserve a double-Weyl point along $\Gamma X$~\cite{Tsirkin:2017lfh}.  Therefore, upon introducing SOC, each spinless double-Weyl point will either split into two Weyl points along the screw line, or, more generically and as occurs in our model, two Weyl points in either of the $k_{y,z}=0$ planes.  Most interestingly, as the spinless double-Weyl points are filling-enforced~\cite{WPVZ,WiederLayers}, then the conventional Weyl fermions in the $k_{y}=0$ plane of our model, while not characterized by any symmetry eigenvalues, are \emph{also} filling-enforced.  This indicates that the large-SOC phase of Eq.~\eqref{eq: SMsg19nodalsurfaceH} is the rare example of a filling-enforced semimetal whose minimal band connectivity~\cite{WiederLayers,QuantumChemistry} is connected through a \emph{low-symmetry BZ plane}.

\section*{SI G. Kramers-Weyl Nodes in Achiral Space Groups}
\begin{figure*}[h]
\centering
\begin{center}
\includegraphics[width=16cm]{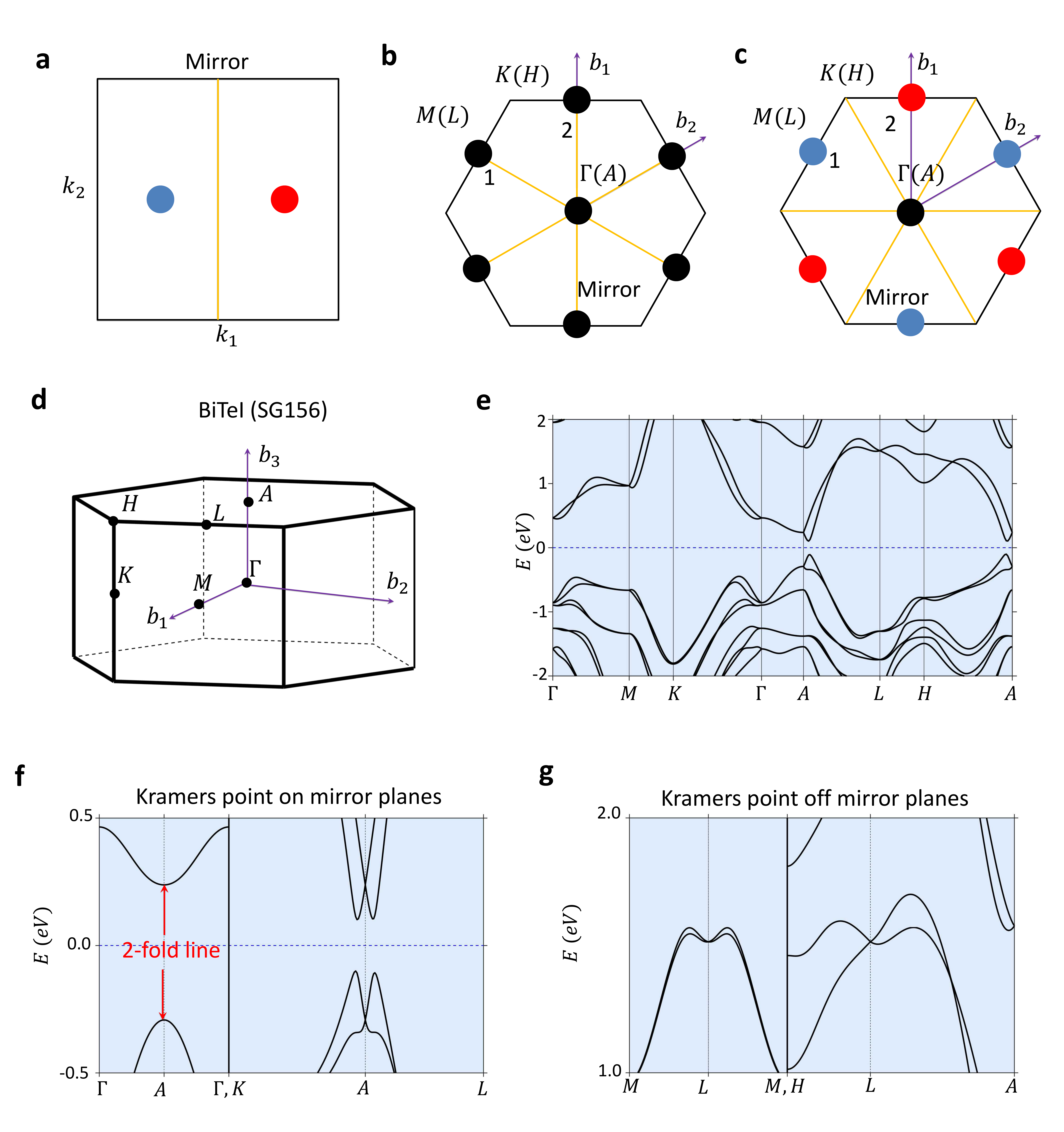}
\end{center}
\caption{\label{Fig2S}\textbf{Kramers points in an achiral group.} (\textbf{a}) A Weyl node and its mirror partner hold opposite chiral charges, which are represented by the blue and red circles. The mirror plane is shown in yellow. (\textbf{b}) A hexagonal lattice whose Kramers points are all on the mirror plane.  In such systems, no Kramers pairs possess chiral charge and, therefore, all are shown in black. (\textbf{c}) A hexagonal lattice with mirror planes that do not intersect its TRIMs exhibits Kramers-Weyl nodes at $L$ and $M$ (\textbf{d,e}). The bulk BZ and electronic band structure of BiTeI (SG 156). (\textbf{f,g}) The band structures in the vicinity of TRIMs for BiTeI.
 }
\end{figure*}

In this section, we discuss the possibility of realizing Kramers-Weyl nodes in achiral space groups with mirror planes. As noted in the main text, any form of roto-inversion is sufficient to remove the geometric ``handedness" of a chiral lattice resulting in a non-chiral (achiral) crystal.   As a Weyl node and its mirror partner have opposite chiral charge, as shown in Fig.~\ref{Fig2S}\textbf{a}, Weyl nodes cannot be stabilized on a mirror plane.  For this reason, Kramers points in achiral space groups, which in most cases are on mirror planes, cannot be chirally charged. To illustrate this, we show in Fig.~\ref{Fig2S}\textbf{b} a hexagonal lattice with mirror planes parallel to unit vectors $b_{1}$ and $b_{2}$, in which we observe topologically trivial Kramers points (black circles) in the BZ. However, if a TRIM does not reside on a mirror plane, then it is still possible for it to realize a Kramers-Weyl node, even in an achiral crystal (as long as inversion symmetry $\mathcal{I}$ is not also present). In Fig.~\ref{Fig2S}\textbf{c} we illustrate a system that has mirror planes not parallel to its unit vectors $b_{1}$ and $b_{2}$. As shown in Fig.~\ref{Fig2S}\textbf{c}, the Kramers pairs at $M$ and $L$ are indeed Kramers-Weyl nodes. Note that $\Gamma$, which is intersected by all the mirror planes, does not possess a Kramers-Weyl node.

To study this case in a real crystal system, we use BiTeI (SG 156) as an example, which is famous for its giant Rashba splitting at its $A$ points. The bulk BZ of BiTeI and its high-symmetry points are shown Fig.~\ref{Fig2S}\textbf{d}, and its electronic band structure along high-symmetry directions is shown in Fig.~\ref{Fig2S}\textbf{e}. Zoomed-in views of the bands near TRIMs $\Gamma$, $A$, $L$, and $M$ are shown in Figs.~\ref{Fig2S}\textbf{f},\textbf{g}, and reveal that bands along $\Gamma-A$ are two-fold degenerate. However, at $M$ and $L$ we observe isolated two-fold degeneracies that are linearly dispersing along all directions in momentum space, and are thus Kramers-Weyl nodes. Conversely, Kramers-Weyl nodes are notably absent at $\Gamma$ and $A$. We note that because the crossings at $M$ and $L$ in BiTeI are too far away from the Fermi level to be enclosed by Fermi surfaces, the Kramers-Weyl nodes at $M$ and $L$ have no effect on the low energy physics.

\section*{SI H.  Fermi Surface Analysis of Kramers-Weyl Materials}

\begin{figure*}[t]
\includegraphics[width=16cm]{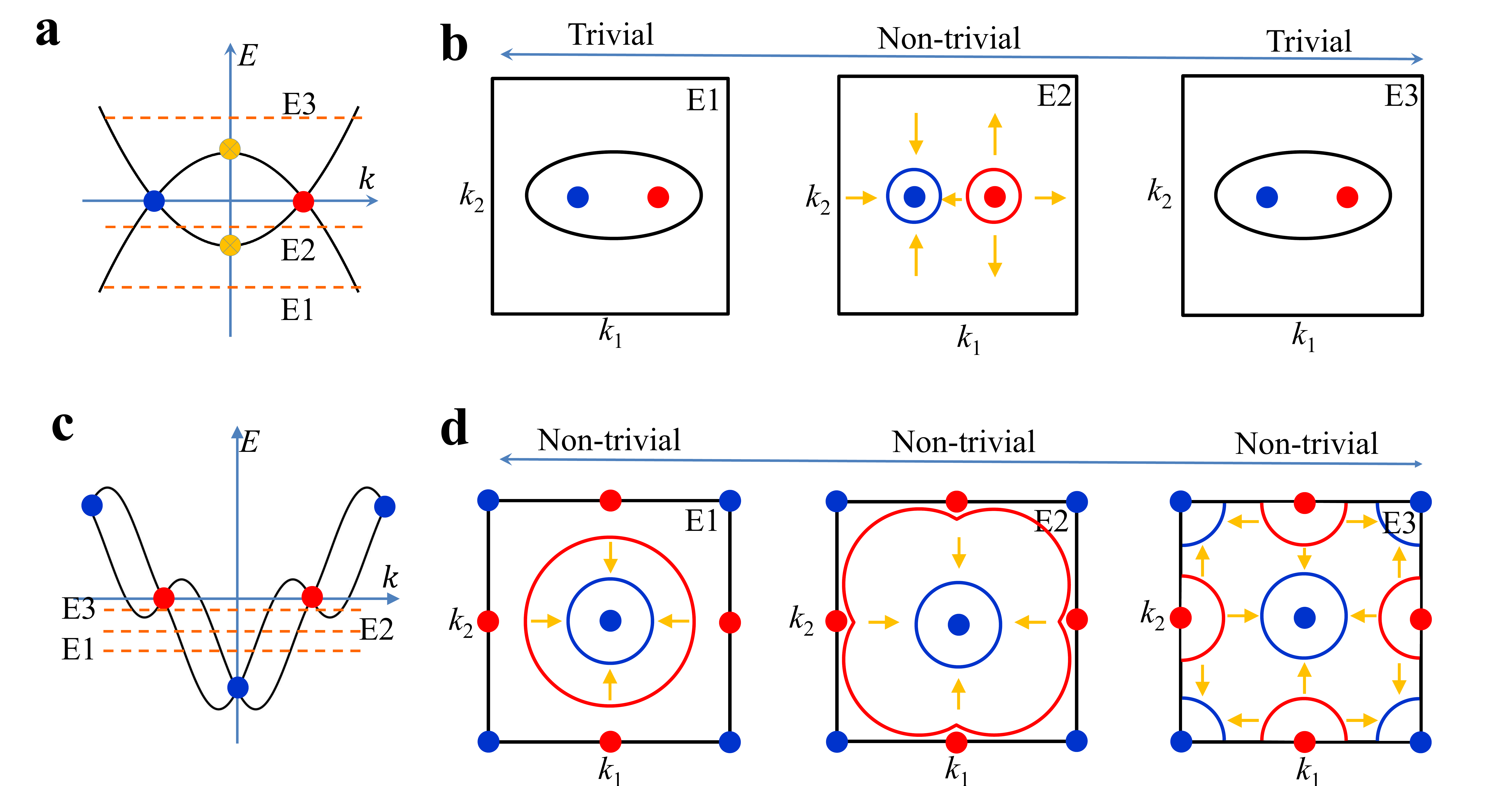}
\caption{{\bf Topological Fermi surface of Kramers-Weyl nodes.} ({\bf a}) The electronic band structure of a conventional band-inversion Weyl semimetal. The blue and red colors represent Weyl nodes of opposite chiral charge. The yellow colors indicate the points where a Lifshitz transition occurs. ({\bf b}) The sequence of Fermi surfaces shown are for energy levels E1, E2, and E3, which are annotated in panel (a). The red and blue coloring of the Fermi surfaces indicates a positive or negative Chern number, respectively. The net chiral charge enclosed by the black Fermi surface is zero and therefore it is topologically trivial. The yellow arrows illustrate the flow of Berry curvature from each topologically nontrivial Fermi surface. The Fermi surface of a band-inversion Weyl semimetal is topologically nontrivial for the energy window defined by the degree of band-inversion, which is typically small. ({\bf c}) The electronic band structure of a typical Kramers-Weyl metal. ({\bf d}) The Fermi surface for energy levels E1, E2, and E3, annotated in panel (c). The Fermi surface of a Kramers-Weyl metal remains topologically nontrivial from the bottom to the top of the bands. Thus, the energy window corresponding to a topologically nontrivial Fermi surface for a Kramers-Weyl metal can be as large as multiple eV.}
\label{FigS3}
\end{figure*}

Comparing the Fermi surface evolution of Kramers-Weyl metals to band-inversion Weyl semimetals, there is a notable contrast in topological character.  Consider the electronic band structure of a band-inversion Weyl semimetal, shown in Fig.~\ref{FigS3}\textbf{a}, in which the energy levels of interest for a Fermi surface analysis are labeled as $E_1$, $E_2$, and $E_3$ and the critical points that correspond to Lifshitz transitions are marked in yellow. Starting at an energy level that is well above the band-inversion Weyl nodes, $E_3$, we observe that the bulk Fermi surface (whose boundary is defined by the black closed contour) encloses the projection of both oppositely chiral charged Weyl nodes and is, therefore, topologically trivial, as shown in the leftmost panel of Fig.~\ref{FigS3}\textbf{b}. Lowering the energy level past the top critical point moves the Fermi surface through a Lifshitz transition. At energy level $E_2$, which is slightly below the pair of Weyl nodes, we observe that the initial Fermi surface splits into two. Generally, at energies between the two Lifshitz transition points, each projected Weyl node is enclosed by its own Fermi surface and therefore, each individual surface is topologically nontrivial, as shown in middle panel of Fig.~\ref{FigS3}\textbf{b}. The yellow arrows in Fig.~\ref{FigS3}\textbf{b} indicate the direction of Berry flux emanating from each projected Weyl node and nontrivial Fermi surface. As the energy level is further lowered, the Fermi surface undergoes a second Lifshitz transition.  At energy level $E_1$, which is close to $E_3$, a single Fermi surface again encloses both projected oppositely chiral charged Weyl nodes, and is therefore topologically trivial, as shown in the rightmost panel of Fig.~\ref{FigS3}\textbf{b}.

Performing a similar evaluation on the model Kramers-Weyl metal in symmorphic chiral space groups, the Fermi surface evolution is drastically different. The cartoon of the electronic structure of a Kramers-Weyl material is shown in Fig.~\ref{FigS3}\textbf{c}.  Starting at $E_1$, we observe two concentric Fermi surfaces around the projected Kramers-Weyl nodes at the BZ center. Unlike the Fermi surface of a band-inversion Weyl semimetal, we instead here observe two distinct, oppositely chiral charged Fermi surfaces enclosing the projected Kramers-Weyl node at the zone center. As shown in Fig.~\ref{FigS3}~\textbf{d}, the inner Fermi surface is formed by the upper cone of the Kramers-Weyl node, while the outer Fermi surface is formed by the lower cone. As the energy level is increased to $E_2$, the nontrivial outer Fermi surface expands towards the zone boundary.  At $E_3$, each Kramers-Weyl node has its own individual nontrivial Fermi surface. The topological properties of these Kramers-Weyl Fermi surfaces therefore differ strongly from those observed or predicted in previous band-inversion Weyl semimetals.  Furthermore, the energy window corresponding to the presence of topologically nontrivial Fermi surfaces in Kramers-Weyl metals, mainly determined by the bandwidth in the limit of vanishing SOC, can be in the energy range of multiple eV.  This property makes experimental observation of topological surface physics in Kramers-Weyl metals promisingly plausible.

\section*{SI I. Fermi Arcs in Kramers-Weyl Metals}

The unique bulk Fermi pockets of Kramers-Weyl Fermions indicate that the appearance of surface states in Kramers-Weyl metals also differs significantly from in band-inversion Weyl semimetals.  In this section, we show the distinct distribution of topological Fermi arcs in a Kramers-Weyl metal, using our tight-binding model of SG-16. In Fig.~\ref{FigARC}\textbf{a} we show the distribution of Kramers-Weyl nodes at the bulk TRIMs. To study the emerging Fermi arc surface states, we consider the projection of Kramers-Weyl nodes onto the (110)-surface. In Fig.~\ref{FigARC}\textbf{b} we show that on this surface, each TRIM has a projected Kramers-Weyl node with chiral charge $\pm 2$. Pictured in Fig.~\ref{FigARC}\textbf{c}, we first calculate the bulk and (110)-surface states for when the SOC strength $t^{s}$ is smaller than lattice hopping strength $t^{1}$. Examining in Fig.~\ref{FigARC}\textbf{d} a constant energy contour at $E=-0.5$ eV reveals that the majority of the projected Kramers-Weyl nodes are covered by bulk pockets that consequently conceal their Fermi arc surface states. However, the projected Kramers-Weyl nodes at $\overline{X}$ and $\overline{M}$ are not covered by the projected bulk pockets and, as shown in Fig.~\ref{FigARC}\textbf{d}, exhibit connecting Fermi arc surface states. To check if the surface states around $\overline{M}$ are consistent with the expected chiral charge projection, we calculate the energy dispersion around a closed path encircling $\overline{M}$, as displayed in Fig.~\ref{FigARC}\textbf{e}. We observe two chiral surface states dispersing along the same $\bs{k}$ direction, confirming that the path defined in Fig.~\ref{FigARC}\textbf{d} encloses a projected Kramers-Weyl node of chiral charge $+2$.

To study the effect of adiabatically increasing SOC, we present in Fig.~\ref{FigARC}\textbf{f} the electronic band structure of our SG-16 tight-binding model for $t^{s}$ much larger than $t^{1}$. As discussed earlier, irrespective of SOC strength, the Kramers-Weyl nodes remain permanently pinned to the TRIMs under the preservation of $\mathcal{T}$ symmetry.  We find that the increase of SOC extensively distorts the bulk bands, allowing for Fermi arcs to be observed over a wider section of the projected surface BZ.   Although the Kramers-Weyl nodes are fixed in momentum space, they are still free to move in energy when SOC is tuned. The resulting constant energy contour at E $=-0.5$ eV for the (110) surface is shown in Fig.~\ref{FigARC}\textbf{g}. Under high-SOC conditions, the projections of Kramers-Weyl nodes are no longer concealed by the projections of bulk pockets and the Fermi arc connectivity becomes easily resolvable. Proceeding as before, the energy-dispersion calculation around a path encircling $\overline{M}$ reveals two co-propagating chiral modes, and again confirms that the enclosed projected Kramers-Weyl node has a $+2$ chiral charge. In general, all of the TRIMs in symmorphic chiral crystals are guaranteed to host Kramers-Weyl nodes, and therefore these systems, if SOC is very large, may host a novel topological Weyl metallic phase with large and well-separated Fermi arcs, similar to the unconventional chiral semimetallic phase predicted in $\beta$-RhSi~\cite{RhSi}.

\begin{figure}[t]
\includegraphics[width=16cm]{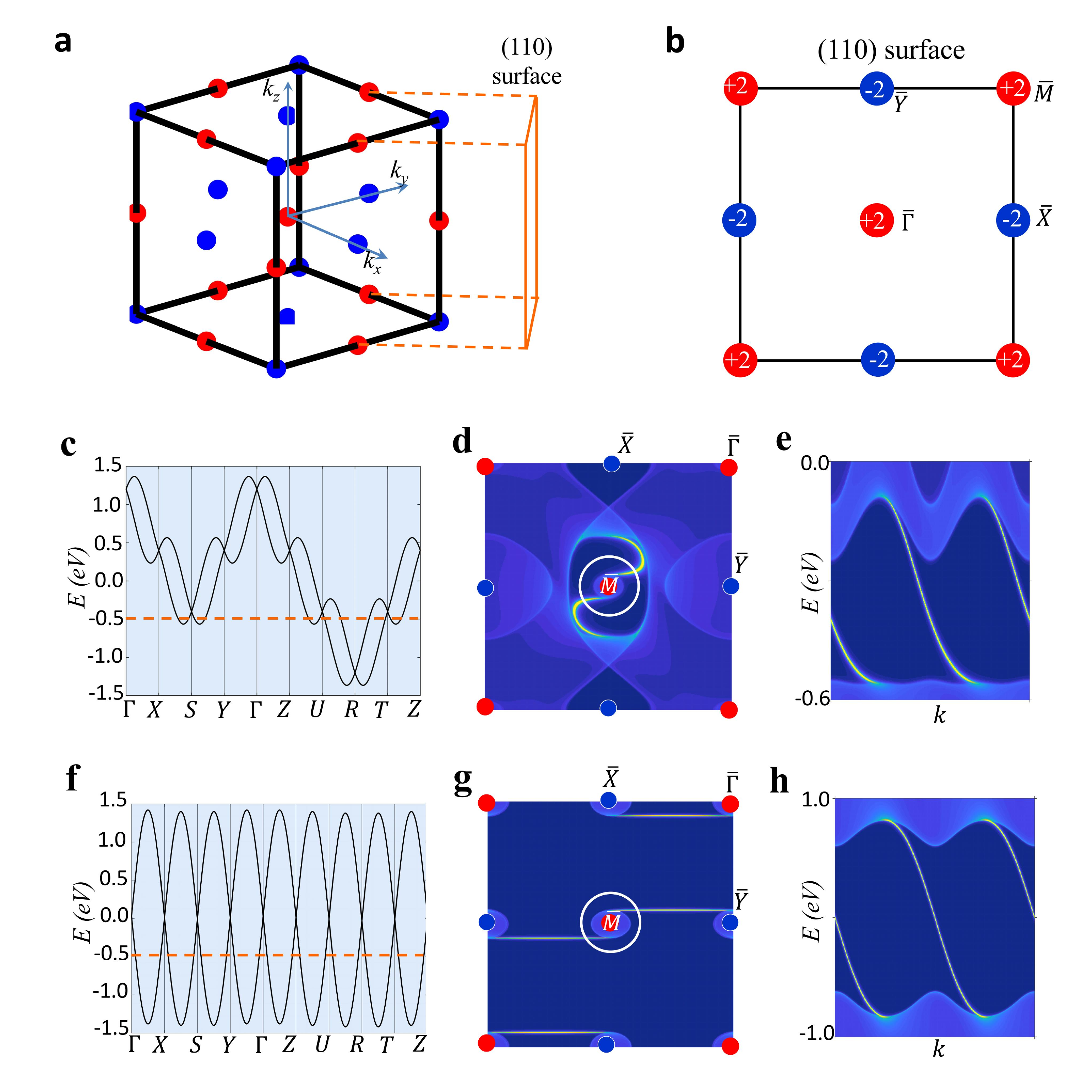}
\caption{{\bf Fermi arcs in a Kramers-Weyl metal.} ({\bf a}) The chiral charge distribution of Kramers-Weyl nodes in the bulk BZ of SG-16. ({\bf b}) The projected chiral charge on the (110) surface. The number labeled in each red and blue marking represents the projected chiral charge at that point. ({\bf c}) The electronic band structure of a Kramers-Weyl metal in SG-16 from the tight-binding model presented in this paper, set with t$_{s}$ smaller or of the same order as t$_{l}$ ({\bf d}) The surface states of the (110) surface at -0.5 eV. Indicated in panel (c) by the orange-dashed line is the energy-level corresponding to (d). ({\bf e}) Energy-dispersion calculation of surface states along the path defined by the white circles enclosing the $\overline{M}$-point. The number of chiral surface states is consistent with the projected chiral charge. }
\label{FigARC}
\end{figure}

\addtocounter{figure}{-1}
\begin{figure*}[t!]
\caption{({\bf f}) We adiabatically change the electronic band structure of the Kramers-Weyl metal by increasing the SOC interaction term t$_{s}$ relative to the lattice hopping term t$_{l}$. Panel (f) shows the band structure when t$_{s}$ is much larger than t$_{l}$. ({\bf g, h}) The surface states calculations from the bands in panel (f) at -0.5 eV.}
\label{FigARC}
\end{figure*}

\section*{SI J. Additional Kramers-Weyl Materials}

\begin{figure*}[t]
\includegraphics[width=16cm]{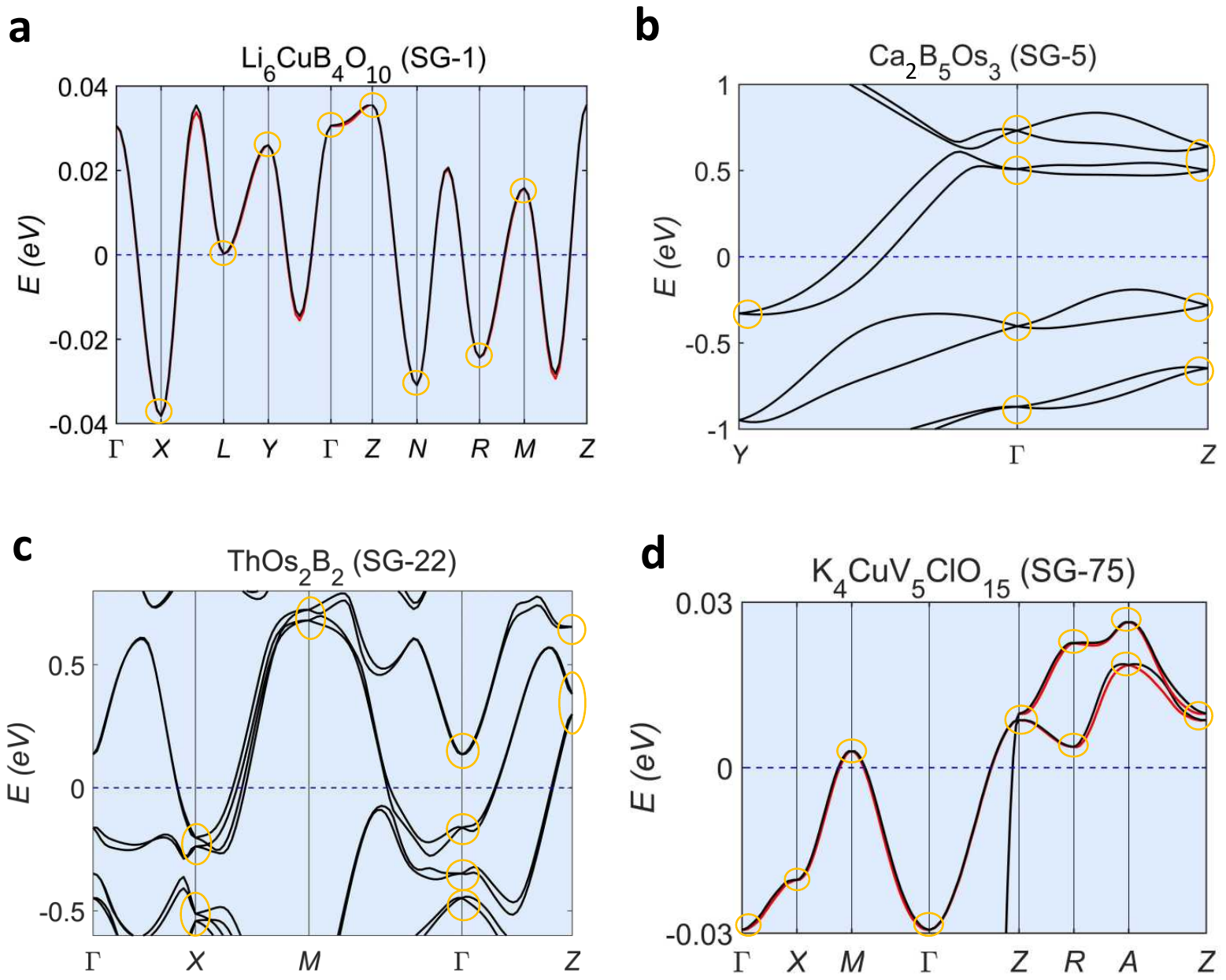}
\caption{{\bf Candidate Kramers-Weyl metals in symmorphic chiral space groups.}}
\label{FigS4}
\end{figure*}

\begin{figure*}[t]
\includegraphics[width=16cm]{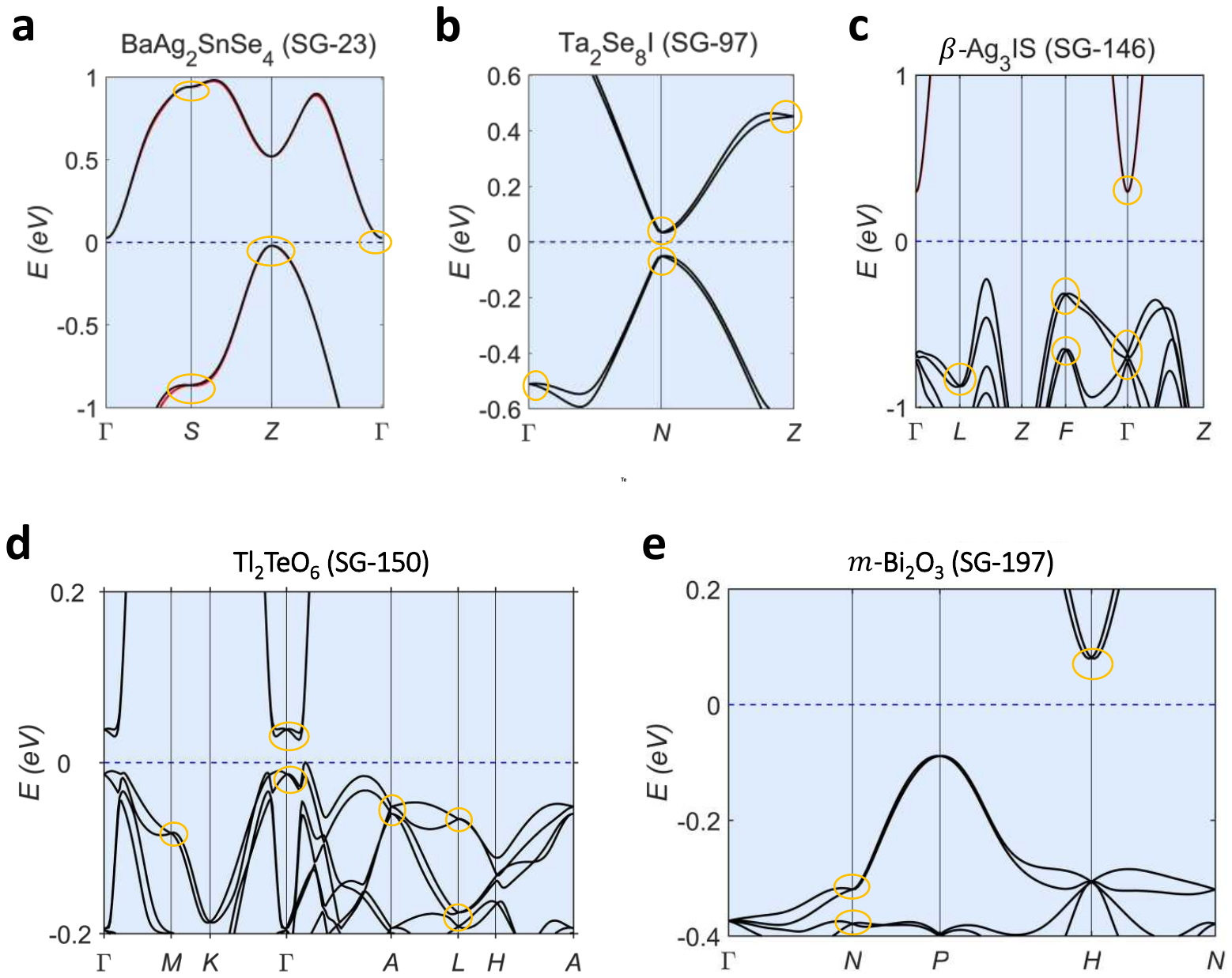}
\caption{{\bf Small-gap insulators in symmorphic chiral space groups with Kramers-Weyl fermions near their Fermi levels.}}
\label{FigS5}
\end{figure*}

\begin{figure*}[t]
\includegraphics[width=16cm]{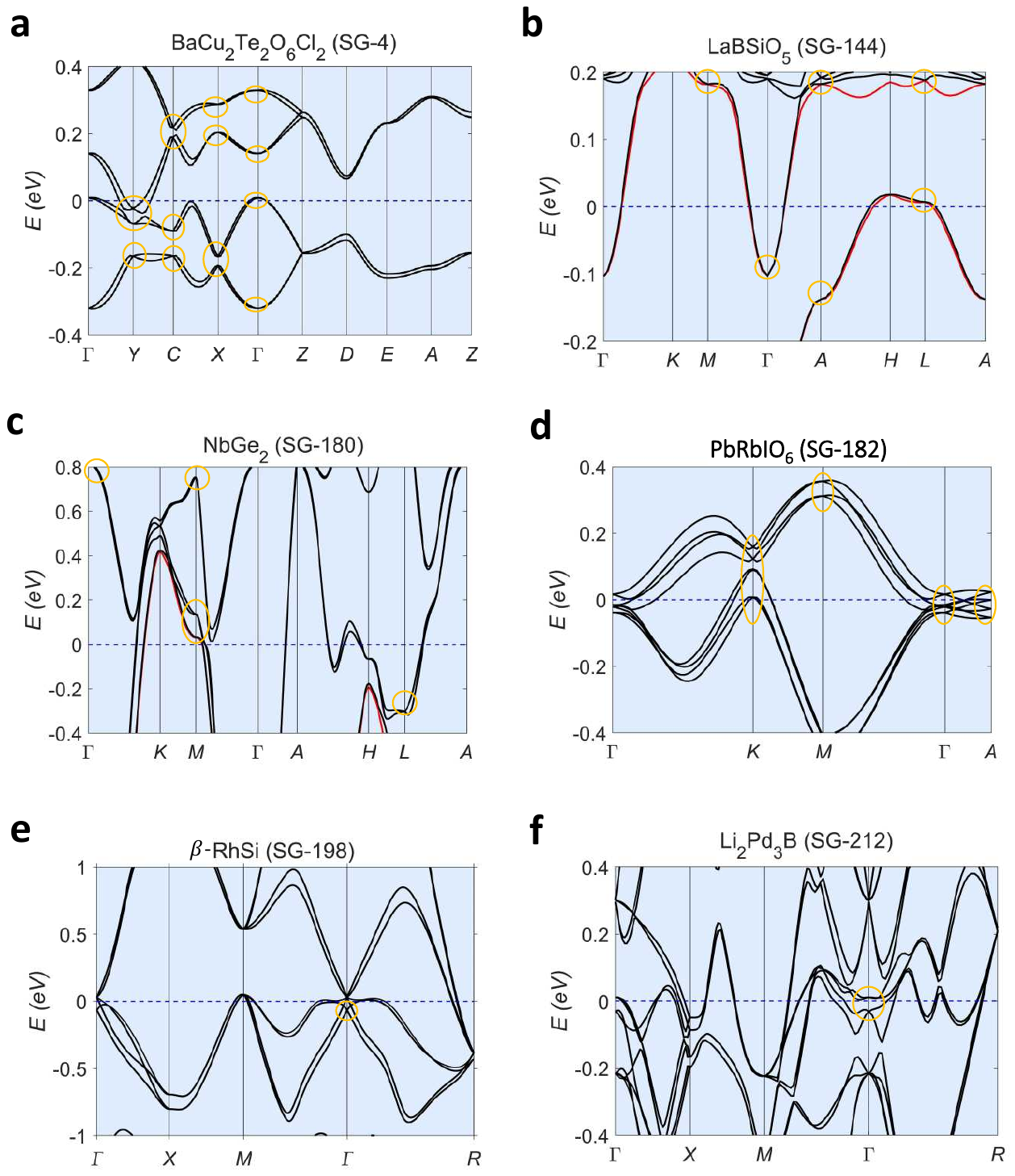}
\caption{{\bf Candidate Kramers-Weyl metals in nonsymmorphic chiral space groups.}  We note that $\beta$-RhSi is also a chiral unconventional fermion semimetal~\cite{RhSi}.}
\label{FigS6}
\end{figure*}

\begin{figure*}[t]
\includegraphics[width=16cm]{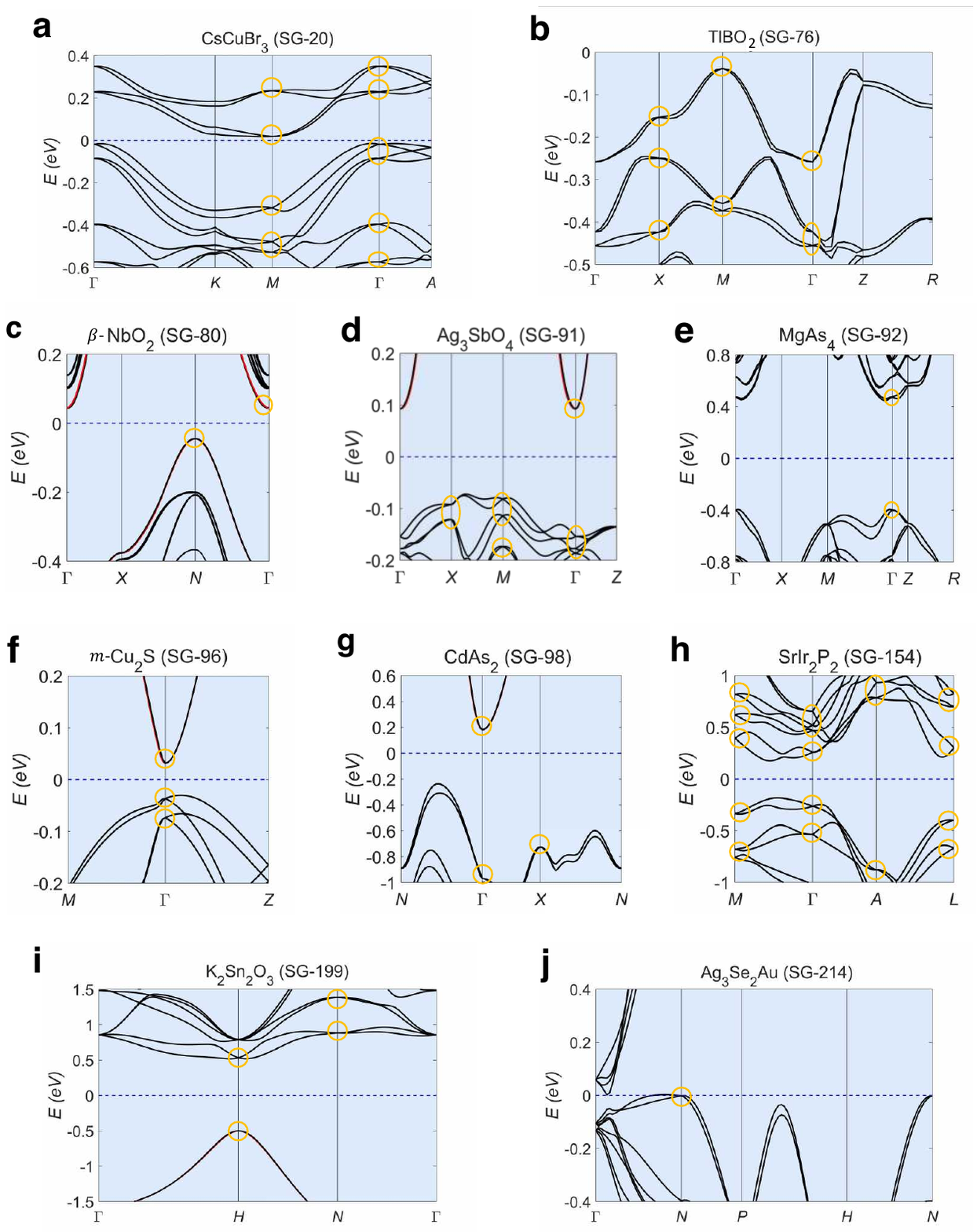}
\caption{{\bf Small-gap insulators in nonsymmorphic chiral space groups with Kramers-Weyl fermions near their Fermi levels.}}
\label{FigS7}
\end{figure*}

Here, we present in Figs.~\ref{FigS4}-\ref{FigS7} additional material candidates that feature Kramers-Weyl fermions.  These materials are either candidate Kramers-Weyl metals with clean Fermi surfaces (Figs.~\ref{FigS4} and~\ref{FigS6}), or are small-gap insulators with Kramers-Weyl fermions near their Fermi levels (Figs.~\ref{FigS5} and~\ref{FigS7}).  Among them, We identify 16 additional chiral materials platforms for the observation of Kramers-Weyl-enabled quantized photocurrent and the chiral and gyrotropic effects: CsCuBr$_{3}$(\#10184) in SG-20, BaAg$_2$SnSe$_4$(\#170856) in SG-23, $\beta-$NbO$_2$ (\#35181) in SG-80, Ag$_3$SbO$_4$(\#417675) in SG-91, MgAs$_4$(\#1079) in SG-92, m-Cu$_{2}$S(\#16550) in SG-96, Ta$_{2}$Se$_{8}$I(\#35190) in SG-97, CdAs$_{2}$ (\#16037) in SG-98, $\beta-$Ag$_3$IS(\#93431) in SG-146, Tl$_{2}$TeO$_{6}$(\#4321) in SG-150, SrIr$_{2}$P$_{2}$(\#73531) in SG-154, m-Bi$_{2}$O$_{3}$(\#27152) in SG-197, K$_{2}$Sn$_{2}$O$_{3}$(\#40463) in SG-199, and Ag$_{3}$Se$_{2}$Au(\#171959) in SG-214 are similar chiral crystals with narrow band gaps, such that moderate doping may allow the isolation of a Kramers-Weyl node at the Fermi level; BaCu$_2$Te$_2$O$_6$Cl$_2$ (\#85786) in SG-4 and Ca$_2$B$_5$Os$_3$(\#59229) in SG-5 are semimetallic crystals which each have Fermi pockets that enclose a single Kramers-Weyl node at intrinsic doping.

\clearpage

\clearpage

\clearpage

\clearpage
\section*{SI K.  Weyl Nodes Distribution in AgBi(Cr$_2$O$_7$)$_2$}

In Table~\ref{Weyl}, we list for reference the locations of both the conventional band-inversion and Kramers-Weyl fermions in AgBi(Cr$_2$O$_7$)$_2$ in SG 79, obtained from first-principles calculations.

\begin{center}
\begin{table}
\begin{tabular}{ c  c  c  c  c c}
\hline

Weyl nodes & $k_x$($\pi/a$) & $k_y$($\pi/a$) & $k_z$($\pi/c$) & Chirality & Number\\
\toprule
\toprule

$W_{1}^{K}$ & 0.000 & 0.000 & 0.000 & -1 & 1  \\

$W_{2}^{K}$ & 0.500 & 0.000 & 0.5000 & +1 & 2  \\

$W_{3}^{K}$ & 1.000 & 0.5000 & 0.500 & +1 & 2 \\

$W_{4}^{K}$ & 0.500 & 0.500 & 0.000 & +1 & 1 \\

$W_{5}^{K}$ & 0.500 & 0.500 & 1.000 & +1  & 1\\

$W_{6}^{K}$ & 1.000 & 1.000 & 1.000 & +1 & 1 \\
\toprule
\toprule

$W_{1}^{I}$ & 0.5 & 0.5 & 0.5 & +1 & 2 \\

$W_{2}^{I}$ & 0.3831 & 0.5493 & 0.4214 & -1 & 8 \\
\hline
\end{tabular}
\caption{\label{Weyl} \textbf{Weyl node distribution in the BZ of AgBi(Cr$_2$O$_7$)$_2$.} Kramers-Weyl fermions are labeled as $W_{m}^{K}$, and conventional band-inversion-generated Weyl nodes are noted as  $W_{n}^{I}$}

\end{table}
\end{center}

\section*{SI L. Theory for Novel Spin Texture of Kramers-Weyl fermions}

In this section we characterize the spin texture of Fermi surfaces enclosing Kramers-Weyl fermions in momentum space. We will first treat the case in which there are no other bands except for the two that participate in the crossing at the Kramers-Weyl node. Remember that conventional band-inversion Weyl fermions can be described in terms of a pseudo-spin degree of freedom that involves superpositions of physical spin and orbital degrees of freedom. In contrast, for Kramers-Weyl fermions we define below a limit in which the pseudo-spin coincides with the physical spin and we give symmetry conditions under which the spin and the momentum around a Kramers-Weyl point are rigidly locked to one another.
The most general effective two-band $\bs{k}�\cdot \bs{p}$ Hamiltonian of a Kramers-Weyl fermion (taken as an expansion about any given TRIM point, e.g., $\Gamma$) acts solely on physical spin. It is given by
\begin{equation}
\label{eq: twobandKramersWeylEffectiveHamiltonianInSM}
\mathcal{H}_{\boldsymbol {k�\cdot p}} =\boldsymbol {d_{k} \cdot \sigma}, \qquad d_{j}(\boldsymbol {k})=\sum_{i=x,y,z}k_iA_{i,j},
\end{equation}
where $\bs{\sigma}$ is the vector of physical spin Pauli matrices and $A$ is a real symmetric $3\times 3$ matrix. Note that this form is valid absent spatial symmetries, which further constrain the values of the matrix $A$ and will be considered later in this section. Note also that a $\bs{k}$-linear term that is proportional to the identity matrix is not allowed by time-reversal symmetry. Such terms lead to a `tilting' of Weyl cones on low-symmetry points in the Brillouin zone. Consider now a spherical surface $\Sigma$ in momentum space that encloses the Kramers-Weyl node. The eigenstates for the momenta on the surface organize into two bands that are in general separated by a bandgap everywhere (since the only enforced crossing occurs at $\bs{k}=0$). The Chern number of any two-dimensional Hamiltonian of the form given in Eq.~\eqref{eq: twobandKramersWeylEffectiveHamiltonianInSM}, as parameterized by $\bs{d}_\bs{k}$, $\bs{k} \in \Sigma$, can be written as
\begin{equation}
C = \frac{1}{4\pi} \int_{\Sigma} \mathrm{d} k_x \mathrm{d} k_y \, \frac{1}{| \boldsymbol {d_{k}} |^3} \boldsymbol {d_{k}} \cdot \left(\frac{\partial \boldsymbol {d_{k}}}{\partial k_x} \times \frac{\partial \boldsymbol {d_k}}{\partial k_y} \right).
\end{equation}

This formula evaluates how often the unit vector $\bs{\hat{d}}_\bs{k} \equiv \bs{d}_\bs{k} /|\bs{d}_\bs{k}|$ covers its target space, the sphere $S^2$, as we vary $\bs{k}$ across the Brillouin zone. Without further spatial symmetries, Kramers-Weyl fermions have $C = \pm 1$. In this case, we know that $\bs{\hat{d}}_\bs{k}$ takes every value on $S^2$ at least once. In addition, we know from Eq.~\eqref{eq: twobandKramersWeylEffectiveHamiltonianInSM} that the energetically higher band $\ket{u^+ (\bs{k})}$ will have its spin expectation value $\braket{\sigma}^+ \equiv \braket{u^+ (\bs{k}) | \bs{\sigma} | u^+ (\bs{k})}$ aligned parallel to $\bs{\hat{d}}_\bs{k}$ at every momentum, while the expectation value $\braket{\sigma}^-$ in the lower band $\ket{u^- (\bs{k})}$ is always antiparallel to $\bs{\hat{d}}_\bs{k}$. This is just due to the fact that in these states, which are by definition eigenstates of the Hamiltonian, we have
\begin{equation}
\braket{\mathcal{H}_{\boldsymbol {k\cdot p}}}^{\pm} =  \boldsymbol {d_k} \cdot \braket{\boldsymbol {\sigma}}^\pm,
\end{equation}
and we know that the states $\ket{u^\pm (\bs{k})}$ maximize (minimize) the expectation value $\braket{\mathcal{H}_{\bs{k}�\cdot \bs{p}}}$. We can thus conclude that within this two-band model for Kramers-Weyl fermions, the expectation value of physical spin sweeps out the entire unit sphere $S^2$ on Fermi surfaces enclosing the Weyl node. However, for conventional band-inversion Weyl fermions, the spin only sweeps parts of the unit sphere $S^2$.

\begin{figure*}[t]
\includegraphics[width=16cm]{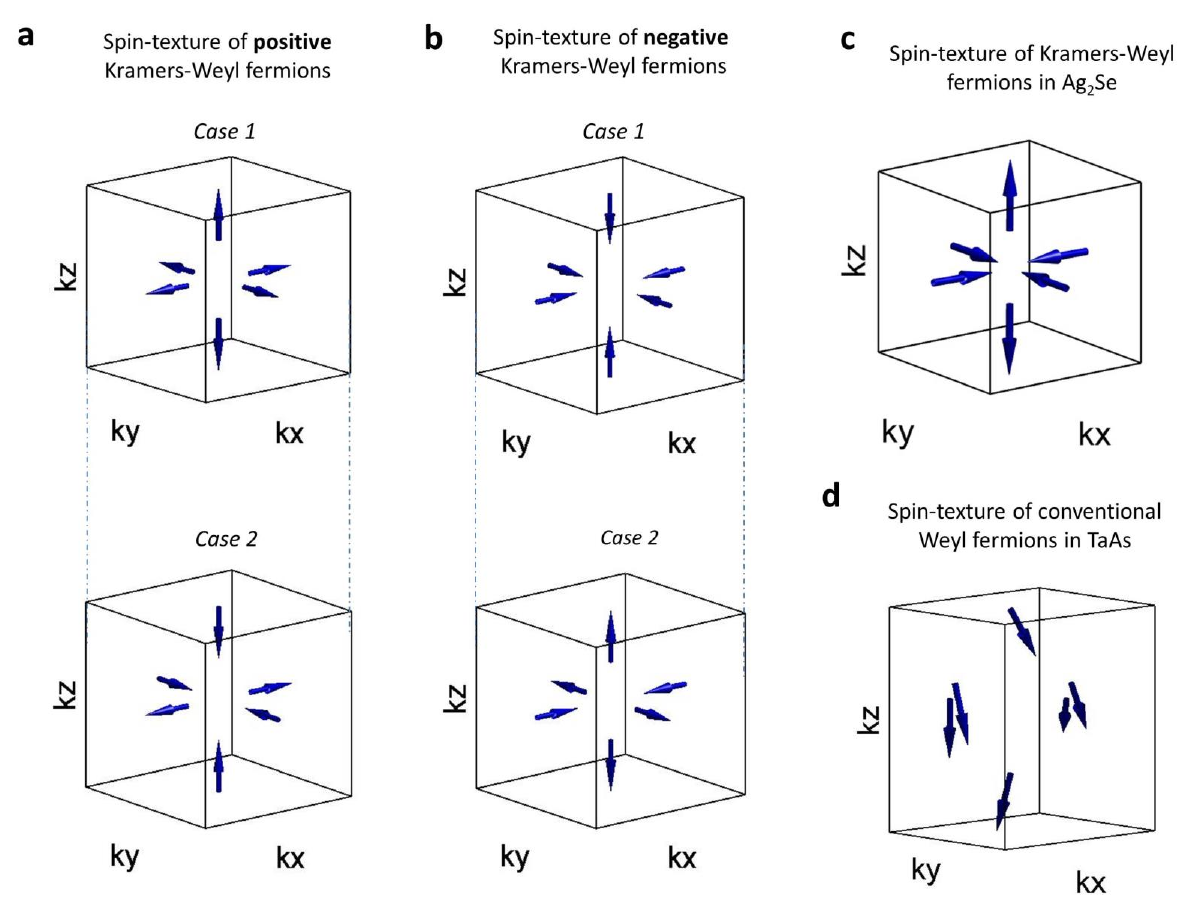}
\caption{{\bf Spin-momentum locking of Kramers-Weyl fermions in the presence of rotational symmetries.} ({\bf a,b}) The spin textures of Kramers-Weyl nodes with opposite chiral charges ({\bf c}) The spin texture of Kramers-Weyl node in Ag$_{2}$Se near its Fermi level. ({\bf d}) The sin texture of a conventional Weyl node in TaAs  }
\label{FigS8}
\end{figure*}

In the presence of rotational symmetries, which can be two-, three-, four- or six-fold in a crystal, we have spin-momentum locking in addition: consider without loss of generality the $z$ direction as rotational axis. Then for $\bs{k} = k_z \hat{\bs{e}}_z$, $\braket{u (\bs{k}) | \sigma_i | u (\bs{k})}$ is only nonzero for $i = z$, since any other direction of the spin expectation value, which would not be parallel to the $z$-axis, would break the rotational symmetry. Therefore, in the case of three mutually orthogonal rotation axes, Kramers-Weyl fermions display perfect spin-momentum locking along these axes.  The chiral point groups with this property all contain $222$, and are therefore $222$, $422$, $622$, $23$, and $432$ (SI \textbf{B}).  In this case, the Chern number can be described by even more straightforward properties of the spin texture. Specifically for a $C$ = +1 Kramers-Weyl node, the spin either points outward along all three principal directions or it points outward along one direction but inward along the other two directions; for a $C$ = -1 Kramers-Weyl node, the spin either points inward along all three principal directions or it points inward along one direction but outward along the other two directions(Figs.~\ref{FigS8}~\textbf{a} and~\textbf{b}).  Note that this is different from conventional Weyl fermions which appear on low-symmetry points in the Brillouin zone away from any rotational axis(Fig.~\ref{FigS8}~\textbf{c} and~\textbf{d}).

Note that this argument holds strictly only in the case of two bands. In a real electronic structure, the bands that cross at a Kramers-Weyl node are however never completely energetically isolated from other bands (which themselves form Kramers-Weyl nodes at the TRIMs). Therefore, hybridization can in principle occur and destroy the pure spin character of Fermi surfaces enclosing the Weyl node. In particular, let $m$ be the separation in energy of a particular Kramers-Weyl node and the energetically closest next node (which may lie lower or higher in absolute energy). Then, the deviation of the actual spin texture from the ideal two-band case described above (for a Fermi surface enclosing the Weyl node with a sufficiently small radius) is small as long as $m \gg c$, where $c$ is the strength of SOC.  This implies that Kramers-Weyl nodes originating from two-dimensional \emph{spinless} corepresentations (\emph{i.e.}, fourfold degeneracies including spin, with vanishing SOC) will necessarily have strong interband coupling and cannot realize this pure spin character.  Specifically, when SOC is reintroduced, these spinless nodes will split in energy, but only with a splitting on the scale of the SOC $m\approx c$.  These higher-dimensional spinless degeneracies comprise the set of weak-SOC corepresentations of the chiral point groups that have complex rotation eigenvalues; they are enumerated in SI \textbf{C}.

\begin{figure*}[t]
\includegraphics[width=16cm]{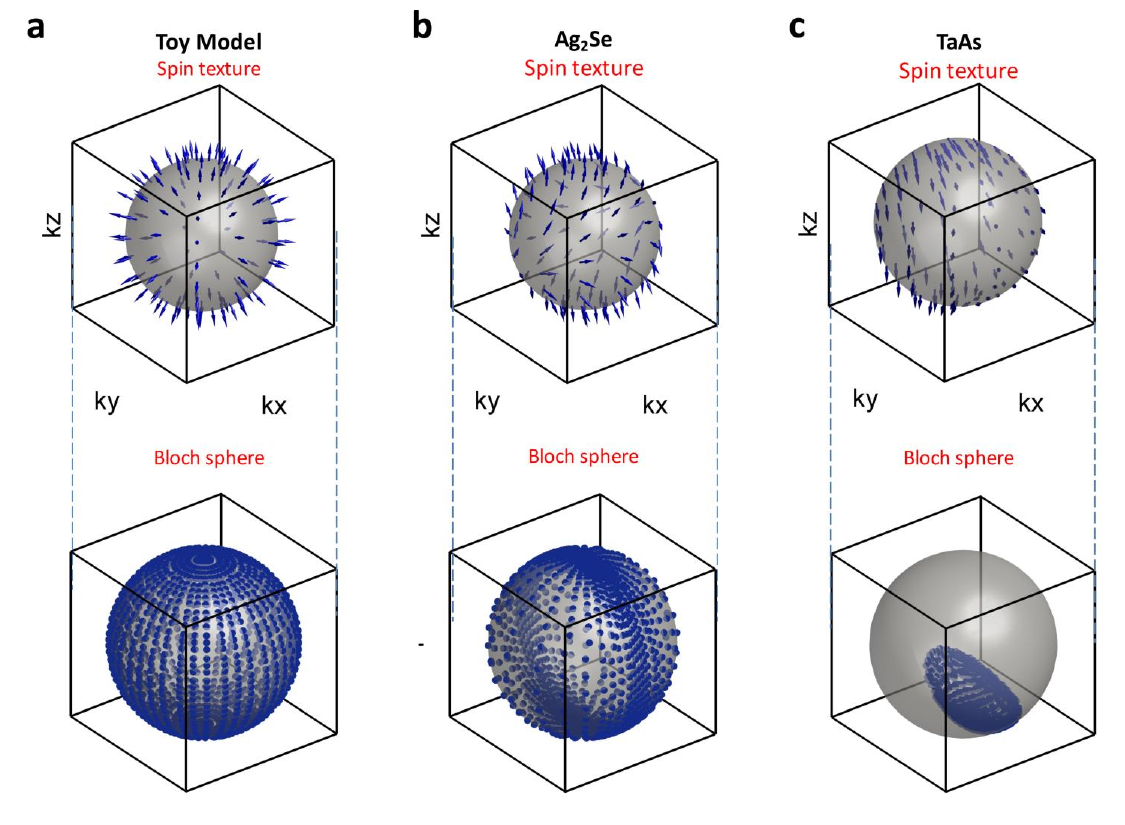}
\caption{{\bf The spin textures of Kramers-Weyl fermions.} ({\bf a}) {\bf Top:} $k$ space spin texture of the Kramers-Weyl fermion in a two-band toy-model. {\bf Bottom:} The projection from the $k$ space spin texture to a Bloch sphere. ({\bf b,c}) The same as panel (a) but for Kramers-Weyl fermion in Ag$_{2}$Se and a conventional band-inversion induced Weyl fermion in TaAs}
\label{FigS9}
\end{figure*}

To demonstrate our theory, we show the spin texture of a $C$=+1 Kramers-Weyl fermion in Ag2Se (Figs.~\ref{FigS8} and~\ref{FigS9}). The physical spin of the Kramers-Weyl fermion in Ag$_{2}$Se indeed sweeps out the full Bloch sphere. Further, because Ag$_{2}$Se is in SG 19 ($C_2$ rotations along $k_{x}$, $k_{y}$, $k_{z}$ axes), spin-momentum locking is observed along all three principle axes, $\boldsymbol {S}_{(k,0,0)} \parallel \boldsymbol {x}$, $\boldsymbol {S}_{(0,k,0)} \parallel \bs{y}$, $\boldsymbol {S}_{(0,0,k)} \parallel \boldsymbol {z}$, where $\boldsymbol {S}_{\bs{k}}$ is the spin expectation value of the occupied band at momentum $\bs{k}$. From the sign of $\boldsymbol {S}_{(k,0,0)} \cdot \boldsymbol {x}$ etc., we can thus compute the Chern number $C^{\mathrm{Kramers-Weyl}} = -\mathrm{sign}(k)\mathrm{sign}(\bs{S}_{(k,0,0)}\times \bs{S}_{(0,k,0)} \cdot \bs{S}_{(0,0,k)} )= +1$. By contrast, the spin of the regular Weyl fermions in TaAs does not sweep out the entire unit sphere, does not show spin momentum locking, and does not correspond to its Chern number.

Therefore, the spin of Kramers-Weyl fermion (1) must sweep out the entire unit sphere; (2) show spin-momentum locking in the presence of additional rotational symmetry; and (3) can be used to measure the Chern number $C^{\mathrm{Kramers-Weyl}}  = -\mathrm{sign}(k) \mathrm{sign}(\bs{S}_{(k,0,0)}\times \bs{S}_{(0,k,0)} \cdot \bs{S}_{(0,0,k)} )$. These are unique properties of the Kramers-Weyl fermions which are absent for a regular band inversion Weyl semimetal.

\clearpage

\section*{SI M.  Calculated Spin Polarization near Kramers-Weyl Points}

To demonstrate how the magnitude of the spin polarization but not the orientation is affected by inter-band couplings in Kramers-Weyl semimetals, consider the following model which pertains to bands at the $\Gamma$ point of space group 16 or 19, for example
\begin{equation}
\mathcal{H}(\bs{k})=
\displaystyle\sum_{i=x,y,z} (v_{i}^{(0)} k_{i} \tau^0\sigma^{i}+v_{i}^{(3)} k_{i} \tau^3\sigma^{i})
+
m\tau^z\sigma^0
+
c\tau^y\sigma^z,
\label{eq: Ham 4-band model}
\end{equation}
where the Pauli matrices $\sigma^x$, $\sigma^y$, $\sigma^z$ as well as the identity matrix $\sigma^0$ act on the physical spin degree of freedom while the Pauli matrices $\tau^x$, $\tau^y$, $\tau^z$ as well as the identity matrix $\tau^0$ act on an orbital degree of freedom, namely a doublet of an $s$ and a $p_z$ orbital. The model respects $C_2$ symmetries around all three coordinate axes given by
\begin{equation}
\begin{split}
&C_2^x\mathcal{H}(k_x,k_y,k_z) \left(C_2^x\right)^{-1}= \mathcal{H}(k_x,-k_y,-k_z),\\
&C_2^y\mathcal{H}(k_x,k_y,k_z) \left(C_2^y\right)^{-1}= \mathcal{H}(-k_x,k_y,-k_z),\\
&C_2^z\mathcal{H}(k_x,k_y,k_z) \left(C_2^z\right)^{-1}= \mathcal{H}(-k_x,-k_y,k_z),\\
\end{split}
\end{equation}
with
$C_2^x=\tau^z\sigma^x$,
$C_2^y=\tau^z\sigma^y$, and
$C_2^z=\tau^0\sigma^z$.
In absence of spin-orbit coupling, only the parameter $m$ in $\mathcal{H}(\bs{k})$ is nonzero and represents the energetic separation between the two orbital/bands. We want to study the limit where spin-orbit coupling is small compared to this parameter, i.e., $c\ll m$ and $\bs{v}^{(0)}\cdot \bs{k}\ll m$, $\bs{v}^{(3)}\cdot \bs{k}\ll m$.

At $\bs{k}=0$, Hamiltonian~\eqref{eq: Ham 4-band model} has two distinct eigenvalues, each of which is two-fold degenerate. These two-fold degenerate subspaces are spilt linearly in $\bs{k}$ at small finite $\bs{k}$.
The possible spin polarization in each subspace follows from the spectrum of the spin operators $\tau^0\sigma^x$, $\tau^0\sigma^y$, and $\tau^0\sigma^z$.  They are given by
\begin{equation}
\begin{split}
\tau^0\sigma^x:&\qquad \pm m/\sqrt{m^2+c^2},\\
\tau^0\sigma^y:&\qquad \pm m/\sqrt{m^2+c^2},\\
\tau^0\sigma^z:&\qquad \pm 1.
\end{split}
\end{equation}
We thus observe that the possible maximal spin polarization of these states (as they become nondegenerate at small finite $\bs{k}$) is reduced from 1 in the $x$ and $y$ direction due to the inter-band coupling $c$.
Thus, if $m\gg c$, i.e., the bands are well separated on the scale of spin-orbit coupling, the spin polarization remains nearly perfect. This conclusion is independent of the presence of the rotational symmetries introduced above.

\begin{figure*}[t]
\includegraphics[width=16cm]{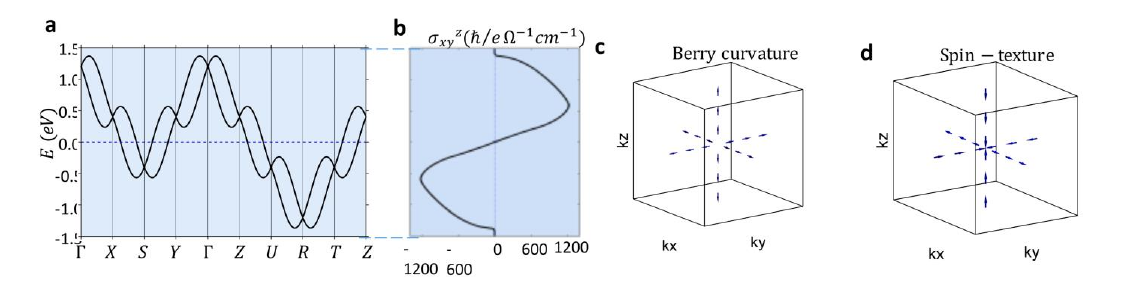}
\caption{{\bf Spin Hall conductivity of Kramers-Weyl fermions}}
\label{FigS10}
\end{figure*}

We also note that the direction (not the magnitude) of the spin polarization at finite $\bs{k}$ away from TRIM points is constrained by the two-fold rotation symmetries. Let $|k_x,k_y,k_z\rangle$ be a nondegenerate eigenstate of the Hamiltonian. Then, for instance,
\begin{equation}
\langle k_x,0,0|\tau^0\sigma^y|k_x,0,0\rangle
=
\langle k_x,0,0|C_2^x\tau^0\sigma^y(C_2^x)^{-1}|k_x,0,0\rangle
=
-\langle k_x,0,0|\tau^0\sigma^y|k_x,0,0\rangle
=
0,
\end{equation}
and by the same argument $\langle k_x,0,0|\tau^0\sigma^z|k_x,0,0\rangle=0$.
Thus, the states along the $k_x$ axis in the Brillouin zone can only have spin polarization in the $x$ direction in spin space, those along the $k_y$ axis can only have $y$ polarization, and those along the $k_z$ axis can only have $z$ polarization. The only remaining freedom is whether the spin polarization in a given nondegenerate band is parallel or antiparallel to the coordinate axis. The spin structure on a spherical Fermi surface around the Weyl point can then take two qualitatively distinct forms (up to permutation of coordinate axes and global spin flips): Either the spin polarization is (anti-)parallel along all coordinate axes, or it has the opposite orientation along one axis with respect to the other two (e.g., parallel along one axis and antiparallel along the other two). Notice that the distinction between these two cases is not in one-to-one relation with the Chern number of the Fermi surface, since flipping the polarization along two coordinate axes (implemented by changing the sign of two of the $v_i$, $i=x,y,z$ parameters) does not change. To reach this conclusion, the presence of the three rotational symmetries introduced above is crucial.

Moreover, the real spin-momentum locking can lead to very large spin Hall conductivity. Specifically, the spin Hall conductivity tensor is expressed as
\begin{align}
\sigma^{\mathrm{Spin Hall}}_{\alpha \beta \gamma} &= \int d^{n} \boldsymbol {k} f(\boldsymbol {k})  \Omega_{\beta \gamma}^\alpha \nonumber\\
\Omega_{\beta \gamma}^\alpha &=-2Im\sum_{n' \neq n} \frac{\langle n \arrowvert \hat{J}^\alpha_\beta \arrowvert n' \rangle \langle n \arrowvert \hat{v}_\gamma \arrowvert n' \rangle}{(E_n-E_{n'})^2},
\label{eq: SpinHall}
\end{align} where $\alpha$, $\beta$, $\gamma$ can be any spatial direction, $f(\bs{k})$ is the Fermi-Dirac distribution function, $\Omega_{\beta \gamma}^\alpha$ is the spin Berry curvature, $\hat{J}^\alpha_\beta=\frac{1}{2}\{\hat{v}_\beta,\hat{s}_\alpha\}$, $\hat{s}$ and $\hat{v}$ are the spin and the velocity operators, $\arrowvert n' \rangle$ is the $n^{\textrm{th}}$ eigenvector. We calculate the a specific component of the spin Hall conductivity, $\sigma^{\mathrm{Spin Hall}}_{xyz}$ for the two-band Kramers-Weyl fermion model in the chiral space group 195 (P23). Fig.~\ref{FigS10} shows the calculated $\sigma^{\mathrm{Spin Hall}}_{xyz}$ can be 1000 $\frac{ \hbar}{e} \Omega^{-1} cm^{-1}$. This is comparable to the value of the largest spin Hall materials such as the 5d transition metals (e.g., Pt). However, it is reasonable to presume that the charge conductivity is smaller than that of the metal Pt because of the smaller carrier numbers. Thus, we expect that the spin Hall angle ($\sigma^{\mathrm{Spin Hall}}/\sigma^{\mathrm{charge}}$) can be even larger than that of Pt. In addition, the spin momentum locking suppresses scattering and thus enhances the spin life time and spin diffusion length.

\section*{SI N.  Nonzero Chern Numbers of all Point Degeneracies in Chiral Crystals}

Here, we further explore our finding that all point-like degeneracies in nonmagnetic chiral crystals with relevant SOC necessarily carry nontrivial Chern numbers.  We first use symmetry and topology to place bounds on the set of allowed point degeneracies in these crystals, and then use Chern number calculations from this and previous works to exhaustively demonstrate that all of these points are necessarily topologically nontrivial.

In a generic 3D crystal, point degeneracies may occur at any point in the BZ; one may find them at high-symmetry points, along high-symmetry lines, in planes of the BZ, or at low-symmetry points throughout the 3D BZ.  In Ref.~\cite{QuantumChemistry}, the authors demonstrated that all degeneracies in crystals with relevant SOC are either described by the corepresentations of the 230 $\mathcal{T}$-symmetric space groups, or are conventional $|C|=1$ Weyl points.  This implies that at any point in the BZ where bands are singly degenerate and there are no additional unitary crystal symmetries, the only degeneracies which may form are conventional Weyl fermions~\cite{Weyl2,Wan,Burkov2011}.  As the 65 chiral space groups necessarily do not contain reflection planes (mirror or glide), as reflection is the operation of a twofold improper rotation $M_{i}=\mathcal{I}\times C_{2i}$~\cite{BigBook,WiederLayers}, then unitary symmetries in chiral crystals are only present along high-symmetry lines and at high-symmetry points in the BZ.  Therefore, we need only demonstrate the nontrivial Chern numbers of point-like degeneracies that are either characterized by a two- (or higher-) dimensional corepresentation at a high-symmetry point, or are characterized by the crossing of two different corepresentations along a high-symmetry line.

We begin by enumerating the allowed point-like degeneracies in chiral crystals at high-symmetry BZ points.  Following the analysis performed in Ref.~\cite{manes} for crystals with negligible SOC, the authors of Ref.~\cite{NewF} discovered that several of the high-fold-degenerate unconventional fermions, such as the threefold-degenerate so-called ``spin-1 Weyl points'' at the $P$ point in SG 214 and the double spin-1 Weyl points at the $R$ point in SG 198 $P2_{1}3$, carry nontrivial Chern numbers.  When the results of that work are combined the recent results of Refs.~\cite{QuantumChemistry,DoubleReps,RhSi,CoSi} and our findings in this work, we conclude that all of the point-like degeneracies at high-symmetry points in nonmagnetic chiral crystals with relevant SOC exhibit nontrivial Chern numbers, and are either (double) spin-1 Weyl fermions, chiral spin-3/2 fermions, or Kramers-Weyl fermions.

Along high-symmetry lines in 3D SGs with $\mathcal{T}$-symmetry, bands may only cross to form two-, three-, or fourfold degeneracies~\cite{BigBook,WiederLayers,DoubleReps}.  Of these, the threefold-degenerate crossings require mirror symmetry~\cite{Triple1,Triple2,Triple3}, and thus may only occur in achiral space groups.  Furthermore, it was demonstrated in Refs.~\cite{Tsirkin:2017lfh,Multi.Weyl,doubleWeyl1,doubleWeyl2} that all twofold point degeneracies along high-symmetry lines are either single-, double-, or triple-Weyl fermions.  Therefore, the only remaining possibilities are the fourfold point-like degeneracies allowed along the BZ edges in chiral space groups with two or more orthogonal screw symmetries~\cite{BarryPrivate,BarryPrep}.  By exhaustively considering all of the corepresentations present along zone-edge lines in these SGs~\cite{QuantumChemistry,DoubleReps}, we find that these fourfold degeneracies come in two categories.  Whether they are filling-enforced~\cite{WPVZ,WiederLayers,RhSi} (for example, in SGs 19 $P2_{1}2_{1}2_{1}$ and 198 $P2_{1}3$), or formed from band inversion (for example, in SGs 18 $P2_{1}2_{1}2$ and 90 $P42_{1}2$), they are locally described by $k\cdot p$ theories with either a twofold or a fourfold rotation axis, as well as the combined antiunitary symmetry $\bar{\mathcal{T}}$ of twofold screw rotation and time-reversal, where $\bar{\mathcal{T}}^{2}=-1$ along the zone-edge line.  We find that all of these fourfold degeneracies carry nontrivial Chern numbers.  Fourfold-degenerate chiral fermions with twofold rotation axes appear in our model of SG 19 in SI \textbf{F}, and can also be obtained by stacking and inverting copies of the 2D model in Ref.~\cite{WiederLayers} of layer group 21, which is isomorphic to SG 18 modulo translations.  This model can then also be tuned to a $C_{4}$-symmetric limit to realize fourfold-degenerate chiral fermions with fourfold rotation axes.  A more detailed discussion of these fourfold-degenerate, zone-edge chiral fermions recently appeared in Ref.~\cite{BarryPrep}.

The nodal degeneracies discussed here comprise all of the possible point-like nodal degeneracies in nonmagnetic chiral crystals with relevant SOC; as they all manifestly exhibit nontrivial Chern numbers, \emph{all} point-like degeneracies in these crystals are necessarily topologically nontrivial.

\section*{SI O.  Comparison between Kramers-Weyl Fermions and Other Chiral Fermions}

It is important to highlight that in the context of symmetry and topology, Kramers-Weyl fermions are distinct from previous examples of chiral fermions.  Conventional band-inversion Weyl semimetals~\cite{Weyl2, Wan, Burkov2011,ARPES-TaAs1, ARPES-TaAs2, ARPES-TaAs3} are not guaranteed by general symmetry criteria, but rather form as the result of favorable energetics.  Furthermore, in many of the conventional Weyl semimetals proposed to date, the degree of band inversion is quite small, and is thus subject to the uncertainties of computational and experimental parameters.  Unconventional high-fold chiral fermions~\cite{manes,NewF, RhSi,CoSi} are protected by highly specific combinations of crystal symmetries, and are thus susceptible to perturbations that break the exact crystal symmetries of their SGs~\cite{DDP}.  Kramers-Weyl fermions, conversely, are neither generated by a band inversion, nor are dependent on specific combinations crystal symmetries; they are guaranteed to exist, and to be chiral fermions, merely by the action of $\mathcal{T}$ symmetry on the irreducible representations of the chiral point groups.  Moreover, as $|C|=1$ Weyl fermions, they can only be destroyed through pairwise annihilation, which may only occur under a large magnetic field that moves them far off from the TRIMs, or through the zone-folding effects of commensurate spin- or charge-density waves.  Thus, Kramers-Weyl fermions are much more robust against perturbations than other chiral fermions.


\begin{thebibliography}{99}
\bibitem{note} There is also a separate, mathematical concept of chiral SGs that is distinct from the definition employed in this work, where in those systems, it is the group algebra which possesses a notion of handedness.  These 22 algebraically chiral groups, also known as the ``enantiomorphic'' groups, comprise the subset of the 65 chiral, or Sohncke, SGs, for which the operation of inversion changes not only the structural handedness of the crystal lattice, but also the set of elements in the SG.~\cite{BigBook}.
\end{thebibliography}
\end{document}